\documentclass{aastex}
\usepackage{emulateapj5}
\usepackage{natbib}

\begin{document}
\slugcomment{Accepted by the Astronomical Journal}

\title{$V$ and $R$-band Galaxy Luminosity Functions and 
Low Surface Brightness Galaxies in the Century Survey}

\author{Warren R.\ Brown\altaffilmark{1}, Margaret J.\ Geller, Daniel G.\
Fabricant, and Michael J.\ Kurtz}

\affil{Harvard-Smithsonian Center for Astrophysics, 60 Garden St.,
Cambridge, MA 02138}

\altaffiltext{1}{Visiting Astronomer, Kitt Peak National Observatory,
National Optical Astronomy Observatory, which is operated by the
Association of Universities for Research in Astronomy, Inc. under
cooperative agreement with the National Science Foundation.}

\email{wbrown@cfa.harvard.edu,
mgeller@cfa.harvard.edu, 
dfabricant@cfa.harvard.edu,
and mkurtz@cfa.harvard.edu
}

\shorttitle{Local Galaxy Luminosity Functions}
\shortauthors{Brown et al.}

\begin{abstract}

	We use 64 deg$^2$ of deep $V$ and $R$ CCD images to measure the
local $V$ and $R$ band luminosity function of galaxies.  The $V_0<16.7$
and $R_0<16.2$ redshift samples contain 1255 and 1251 galaxies and are
98.1\% and 98.2\% complete, respectively.  We apply $k$ corrections before
the magnitude selection so that the completeness is to the same depth for
all spectral types.  The $V$ and $R$ faint end slopes are surprisingly
identical: $\alpha=-1.07\pm0.09$. Representative Schechter function
parameters for $H_0=100$ are: $M^*_R=-20.88\pm0.09$,
$\phi^*_R=0.016\pm0.003$ Mpc$^{-3}$ and $M^*_V=-20.23\pm0.09$,
$\phi^*_V=0.020\pm0.003$ Mpc$^{-3}$.  The $V$ and $R$ local luminosity
densities, $j_R=(1.9\pm0.6)\times10^8 L_\sun$ and
$j_V=(2.2\pm0.7)\times10^8 L_\sun$, are in essential agreement with the
recent 2dF and SDSS determinations.

	All low surface brightness (LSB) galaxies fall in the large scale
structure delineated by high surface brightness galaxies.  The properties
and surface number density of our LSB galaxies are consistent with the LSB
galaxy catalog of O'Neil, Bothun \& Cornell, suggesting that our samples
are complete for LSB galaxies to the magnitude limits.  We measure colors,
surface brightnesses, and luminosities for our samples, and find strong
correlations among these galaxy properties.  The color-surface brightness
relation is $(\vr)_0=(-0.11\pm0.05)\mu_{R,0} + (2.6\pm0.9)$.

\end{abstract}


\section{INTRODUCTION}

	The galaxy luminosity function (LF), the number of galaxies per
unit luminosity per unit volume, is a fundamental quantity in
observational cosmology.  The galaxy LF figures prominently in the study
of galaxy evolution, interpretation of faint galaxy counts, and
measurement of the luminosity density of the Universe.  Deep but narrow
``pencil-beam'' redshift surveys, including the Canada France Redshift
Survey \citep{lilly95}, the Autofib Redshift Survey \citep{ellis96}, and
the Canadian Network for Observational Cosmology Redshift Survey
\citep{lin99}, probe evolution of the galaxy LF for $z<1$. The local
galaxy LF is the benchmark for these studies.

	The local galaxy LF is measured from large angular scale redshift
surveys; Table \ref{lfs} summarizes all of these surveys with $>1000$
galaxies.  The LF parameters are given in the \citet{schechter76}
parameterization, where $\phi^*$ is the normalization (galaxies per
Mpc$^{-3}$), $\alpha$ is the faint end slope, and $M^*$ is the
characteristic absolute magnitude at the ``break'' between the exponential
bright end and power-law faint end of the LF.  The Center for Astrophysics
Redshift Survey \citep[CfA2,][]{marzke94} and the complementary Southern
Sky Redshift Survey \citep[SSRS2,][]{marzke98} are complete but relatively
shallow surveys.  The Stromlo-APM Redshift Survey
\citep[APM,][]{loveday92} and Durham/UKST Galaxy Redshift Survey
\citep[D/UKST,][]{ratcliffe98} are sparsely-sampled surveys of modest
depth. The ESO Slice Project Redshift Survey \citep[ESP,][]{zucca97} and
the on-going 2 Degree Field Galaxy Redshift Survey \citep[2dF,][]{cross00}
are both deeper surveys with mean redshift $\overline{z}=0.10$.  Only the
Las Campanas Redshift Survey \citep[LCRS,][]{lin96} and the Sloan Digital
Sky Survey \citep[SDSS,][]{blanton01} are CCD-based, and together with the
Century Survey \citep[CS,][]{geller97} are the only $R$-selected surveys.
Multiple passband photometry has not been available for these complete,
large galaxy surveys except for the SDSS.

	We have deep $V$- and $R$-band CCD images for 64 deg$^2$ of the
Century Survey \citep[CS,][]{geller97}.  The original CS is a 102
square degree redshift survey containing 1762 galaxies to a limiting
magnitude $R=16.13$.  The original CS is complete and requires no sampling
corrections; selection at $R$ minimizes galactic extinction and $k$
corrections.  The original CS LF parameters, based on CCD-calibrated
photographic photometry, are comparable to those derived from surveys of
similar depth (Table \ref{lfs}).

	Here we determine the local galaxy LF by extending the original CS
$R$ sample to $R_0 < 16.2$ mag and present a new $V$ sample to $V_0 <
16.7$ mag.  The redshift samples are 98\% complete and contain $\sim1250$
galaxies.  We give attention to details of the construction of clean
redshift catalogs, and offer the following subtle, but important,
modifications to previous approaches to local LF determinations.

	We have $V$ and $R$ photometry for the same region of sky.  The
SDSS \citep{blanton01} is the only other large, local redshift survey with
multi-passband photometry.  Accurate colors allow us to constrain the $k$
correction for each galaxy.  We apply $k$ corrections {\it before} the
magnitude selection to ensure that we sample different spectral types 
to the same depth.

	The CCD photometry has average errors $\pm0.049$ mag, much lower
than the $\pm0.25$ mag errors in the original photographic photometry. The
CCD images have approximately twice the resolution of scanned photographic
plates, allowing us to better separate close pairs of galaxies.  Our
redshifts, measured from long-slit spectra, are complete even in high
density regions and have no surface brightness constraints.

	One of the major concerns in measuring the galaxy LF is the proper
inclusion of low surface brightness galaxies. \citet{sprayberry97}, for
example, estimate that the CfA redshift survey may miss 1/3 of the local
galaxy population because of the limiting surface brightness of the
photographic photometry.  Our CCD photometry extends to a limiting sky
brightness of 26 $R$ mag arcsec$^{-2}$, allowing a test of the
photographic CS completeness for low surface brightness galaxies.  Our
data also allow an examination of the relative distribution of low surface
brightness galaxies and the large scale structure defined by high surface
brightness galaxies.

	In section 2 we describe the data and errors, and in section 3 we
define the photometric samples. In section 4 we study the characteristics
of the galaxies, emphasizing the properties and distribution of low
surface brightness galaxies. In section 5 we discuss the LF determination.  
We conclude in section 6.  Appendix A contains a detailed comparison of
the photographic and CCD-based CS photometry.  Appendix B presents the the
$V$ and $R$ CCD photometry.

\section{DATA}

	We obtained Johnson $V$ and Kron-Cousins $R$ broadband images with
the MOSAIC camera \citep{muller98} on the Kitt Peak National Observatory
0.9m telescope in 1998 December and 1999 February.  MOSAIC is an 8 CCD
array with pixel scale 0.424 arcsec pixel$^{-1}$ and field of view
$59^\prime $ x $59^\prime $ on the 0.9m. We obtained three offset 200 sec
$R$ and 333 sec $V$ images at each position along the western 64 deg$^2$
of the CS slice \citep{geller97}:  $8\fh5 < \alpha_{\rm B1950} < 13\fh5$,
$29^\circ < \delta_{\rm B1950}< 30^\circ$.

	To complete the sample to $V_0=16.70$ and $R_0=16.20$ mag, we
measured 242 new redshifts with the FAST spectrograph \citep{fabricant98}
on the Fred Lawrence Whipple Observatory 1.5m telescope in 2000 April and
2000 November \citep[to be published in][]{wegner01}.  We used the 300
line grating to obtain 6 \AA\ FWHM resolution and spectral coverage from
3650 \AA\ to 7500 \AA. We obtained velocities with the cross-correlation
package RVSAO \citep{kurtz98}.  The mean uncertainty of the velocities is
$\pm40$ km s$^{-1}$.  The new redshifts make our $V$ and $R$ samples
98.1\% and 98.2\% complete, respectively.

\subsection{Processing}

	We obtained bias and flat field calibration images in the standard
fashion, and processed them with the $IRAF$ MSCRED package.  The MOSAIC
images contain variable row-to-row bias fluctuations of $\pm3$ counts as
well as large scale moir\'{e} patterns of amplitude $\lesssim 3$ counts.  
Drifts in the various clocks in the CCD controllers produce the variable
bias levels; they are not reproducible or completely removable (Valdes
1999, private communication).  To reduce the row-to-row bias fluctuations,
we fit and subtract the overscan region on a row-to-row basis.  This
procedure leaves large-scale bias features near chip edges.  To remove
these, we smooth the median-combined bias images with a 20 pixel box and
subtract them from the data.  This approach avoids introducing additional
noise from the non-repeating moir\'{e} patterns.

	The moir\'{e} patterns modulate the dome flats by only 1 part in
5000; thus we use these flats in the standard fashion.  This procedure
leaves center-to-edge background changes of $\sim10\%$, resulting from the
different colors of the dome flat lamp and the night sky.  ``Night sky
flats'' are a median combination of a night's worth of data images (scaled
to a common background level) for each filter.  We smooth the night sky
flats with a 20 pixel box to increase their S/N. This approach minimizes
the introduction of additional noise at small scales; bad pixel masks
eliminate hot pixels and bad columns from the smoothing.  Division by the
night sky flats flattens the background to 2\% at large scales. Dark
images have means of $\leq 0.1$ counts, and we make no correction for dark
current.

\subsection{Astrometry}

	Good astrometry is critical for working with multi-chip data.  
The MOSAIC optical system causes a 2.0\% pixel area change over the field
of view of the KPNO 0.9m \citep{davis98}.  We re-bin the processed images
with $IRAF$ using the astrometric solutions provided to correct the scale
changes.  We then use WCSTools \citep{mink99} to refine the zero-point and
rotation of the coordinate system for each chip.  We refine the coordinate
system with the Hubble Space Telescope Guide Star Catalog
\citep{lasker90}, excluding the 5 brightest stars per field to cull the
objects with the worst proper-motion.  Depending on seeing conditions and
star density, typical coordinate system fits use $\sim20$ stars per chip
and provide coordinates good to $0\farcs3 - 0\farcs5$.

\subsection{Photometric calibration}

	We use Landolt fields \citep{landolt92} and the M67 cluster
\citep{montgomery93} for flux calibration.  The M67 photometry, although
less precise than the Landolt photometry, extends over a half degree and
is available for 1000+ stars.  The flux calibration is typically based on
1000 $R$ band and 7500 $V$ band standard star measurements per night.

	We performed a series of tests on the flux calibration images.
First, to test for linearity in the CCDs we imaged M67 at low airmass with
back-to-back 10 and 100 second exposures. The difference in stellar
magnitudes between the two exposure times is $\overline{\delta m} =
-0.004\pm0.005$ mag.  We conclude there is no significant non-linearity
over the count rates appropriate for the galaxies.

	We compare magnitude zero points for the \citet{landolt92}
14$\arcsec$ circular aperture magnitudes and the adaptive-aperture
``total'' magnitudes that we use for galaxies.  The zero-point solutions
for both aperture magnitudes are within $1\sigma$ on the nights of best
and worst seeing.  Thus we adopt adaptive-aperture total magnitudes
throughout.  Finally, we compare the photometric solution based on the
Landolt fields alone to the full photometric solution including the M67
fields.  The photometric solutions agree to $\le 1\sigma$, giving us
confidence in the M67 photometric calibrations.

	We compute the photometric solution as follows.  We do not fit the
zero points, extinction coefficients, and color terms simultaneously
because of degeneracy in the photometric solution.  The color term is
small and easily washed out by errors in the zero point and extinction
coefficient; thus we first solve for the color term.  Using standards
taken at identical airmass on the same night (so that the extinction
coefficient is constant), the color term is the slope of the $(m_{\rm
instrumental} -m_{\rm true})$ vs.\ color line.  Only the Landolt field in
chip 2 allows an accurate color term determination (see Figure
\ref{figct}); the small range of (\vr) colors in M67 provide little
leverage on the color term.  We used the chip 2 color term for all 8 chips
because all 8 chips look through the same filter, and because the color
term is primarily affected by filter transmission.  The color term values
for the other 7 chips agree within their errors with the chip 2 value.  
The color terms are constant from night to night; thus we use the average
color term for all 7 nights of observations:  CT$_R=-0.013\pm0.005$ mag
and CT$_V=0.072\pm0.005$ mag.

	Once we subtract the color term from the standard star
observations, we determine the extinction coefficient from a linear fit of
the magnitude vs.\ airmass line.  The extinction coefficient is
atmospheric and we assume it is identical for all 8 chips. The nightly
fluctuations in the extinction coefficient were of order the error; we use
the average extinction coefficient of all nights in an observing run.  
Finally, subtracting the color term and extinction term from the standard
star observations, we determine the zero-point for each individual chip,
for each night.  Depending on the nightly sky conditions, the residual RMS
error in the standard star calibration is $\pm0.02-0.04$ mag.

\subsection{Galaxy photometry}

	We use SExtractor \citep{bertin96} to make $10^4$ photometric
galaxy measurements in our 300 Gb of images.  SExtractor performs
background subtraction, convolves the images with a detection filter, and
measures circular aperture, isophotal, and adaptive-aperture magnitudes.  
The adaptive-aperture magnitudes are elliptical apertures that include
$\simeq90\%$ of the flux and are corrected to total magnitudes assuming
Gaussian convolved profiles.  We use these total magnitudes throughout.

	We set a detection threshold of 25 $R$ mag arcsec$^{-2}$. Our
limiting sky brightness is 26 mag arcsec$^{-2}$ in both passbands. We
apply $R$ apertures to aligned $V$ images. This process assures that the
light in each bandpass comes from the {\it same physical region} of a
galaxy.  With $\sim7000$ objects per image, we can align $V$-band images
to within $\pm0.05$ pixels of their companion $R$-band images.

	We use a gaussian convolution filter with FWHM greater than the
seeing to optimally detect extended galaxies.  SExtractor detects the low
surface brightness galaxies found by visual inspection, including an
extreme object with an exponential scale length of $30\arcsec$ and central
surface brightness 23.4 mag arcsec$^{-2}$ in $V$ (at $12^{\rm h}31^{\rm
m}58\fs42, 29^\circ42\arcmin33\farcs2$ J2000).

	We follow \citet{sprayberry97} and define low surface brightness
(hereafter LSB) galaxies as extended objects with $\mu(0)_B > 22$ mag
arcsec$^{-2}$, where $\mu(0)$ is the central surface brightness
for an exponential profile,
	\begin{equation}
\mu(r)=\mu(0) + 1.086r/\alpha_l.
	\end{equation} $\mu(r)$ is the surface brightness as a function of
radius $r$ and $\alpha_l$ is the scale length in arcseconds.  We take the
average (\bv)=0.79 mag LSB galaxy color from \citet{oneil97b} and the
average (\vr)=0.40 mag from our own sample of LSB galaxies to define LSB
galaxies in the $V$ and $R$-bands as objects with $\mu(0)_V > 21.2$ and
$\mu(0)_R > 20.8$ mag arcsec$^{-2}$.

	We select galaxies from each $V$ and $R$ catalog with a
combination of FWHM, isophotal area, and central surface brightness
criteria.  We make a magnitude cut at $R=16.5 + A_{R {\rm max}}$, and then
visually examine every selected object.  Our photometry recovers all
original CS galaxies.  The CCD imaging resolves 2\% of the original CS
objects into multiple objects (Appendix A).

	Approximately 20\% of the $10^4$ galaxy measurements are either
cut off by CCD boundaries, fall on cosmic rays or bad columns, or are
poorly fit by SExtractor.  The poorly fit galaxies are commonly irregular
LSB galaxies, galaxies near a bright star, or unusually large galaxies.  
We obtain photometry of poorly fit galaxies by hand with a modified
version of GALPHOT \citep{freudling93}, a front end for the $IRAF$ ELLIPSE
package.  This procedure allows us to mask unwanted artifacts and to
center apertures on difficult objects properly.

	SExtractor and GALPHOT total magnitudes appear equally robust.  
We fit surface brightness profiles, using GALPHOT, to 101 randomly
selected CS galaxies.  Comparison of the total integrated magnitudes to
the SExtractor total magnitudes yields a 1$\sigma$ scatter of $\pm0.05$
mag and a -0.01 mag offset.  On average, integrating the GALPHOT profile
yields only 1\% more flux than SExtractor total magnitudes.

	We imaged each square degree at least 3 times.  We compute final
magnitudes and positions from the weighted averages of individual
measurements.  This approach avoids the need to stack the offset images
and takes advantage of the good S/N of the CS galaxies. Using multiple
measurements allows us to measure our errors: Figure \ref{figerr} shows
the distribution of the RMS scatter of identical galaxies measured on
different nights.  The median RMS error is $\pm0.025$ mag; this value
includes both measurement and zero point errors.  When we plot the average
magnitude offset between fields observed on different nights,
non-photometric conditions, including 0.05 mag extinction from cirrus
bands, are strikingly obvious.  We add a corrective offset to catalogs
with cirrus extinction less than 0.2 mag, and assign larger errors to
those catalogs.  We see no structure in the cirrus extinction at the 1
degree scale during our exposures.

	The final photometric error (Appendix B) is a sum in quadrature of
the zero-point error and the RMS photometric scatter determined for
multiple measurements of each galaxy.  The average final photometric error
is $\pm0.049$ mag in both passbands.

	Figure \ref{figcscomp} compares our CCD photometry with the
photographic CS photometry.  The zero-point offset $\Delta
R_{CCD-photographic} = 0.014 \pm 0.22$ mag is small and well within our
average zero-point error $\pm0.034$ mag.  The $\pm0.22$ mag RMS scatter
between the CCD and photographic photometry is consistent with the
$\pm0.25$ mag photographic error estimate of \citet{geller97}.  Appendix A
compares the CCD and photographic catalogs in detail.

\subsubsection{Corrections}

	We take galactic extinction corrections from \citet{schlegel98},
who combine $COBE$/DIRBE and $IRAS$/ISSA data to map dust emission in the
Milky Way.  The CS strip is at high galactic latitude, thus the average
extinction is $A_R=0.053$ and $A_V=0.065$ mag (Figure
\ref{figka}a). \citet{schlegel98} estimate that the errors in the
extinction correction are $\sim16\%$ of the extinction.

	The final photometry also includes $k$ corrections to correct
observed magnitudes to restframe magnitudes.  Because we know the (\vr)
color and the redshift of each galaxy, we can better constrain the $k$
correction for each galaxy.  We interpolate tables provided by
\citet{poggianti97} for types E, Sa, and Sc, and \citet{frei94} for type
Im. The k corrections range $0 < k_R < 0.20$ and $-0.05 < k_V < 0.30$ mag.  
The average values are $k_R=0.061$ and $k_V=0.089$ mag (see Figure
\ref{figka}b).  Figure \ref{figkvr} demonstrates the efficacy of the k
corrections: after correction the apparent cosmological reddening is
almost entirely removed.

	The errors in the $k$ correction are difficult to quantify.  
There is significant spectral variation within morphological classes of
galaxies, but we bypass the usual difficulty of determining morphologies
by using colors.  Broadband colors more tightly fit a given {\em spectral}
type; colors are effectively very low resolution spectra.  We obtain one
estimate of the $k$ correction error by propagating the (\vr) error.  The
errors are of order of the $k$ correction itself.  Thus we make the
conservative assumption that the $k$ correction error is the value of the
$k$ correction.  If we were to ignore the $k$ correction error in the LF
calculation, the faint end slope would be $-0.04$ steeper and the
characteristic absolute magnitude $M^*$ $-0.04$ to $-0.09$ mag brighter
(see Equation \ref{schfn} below).

\section{SAMPLE DEFINITION}

	Our data cover the CS strip $8^{\rm h}32^{\rm m}45^{\rm s} <
\alpha_{\rm B1950} < 13^{\rm h}27^{\rm m}31^{\rm s}$,
$29^\circ0\arcmin0\arcsec < \delta_{\rm B1950} < 30^\circ
0\arcmin0\arcsec$ for a total of 64.13 deg$^2$.  We find that 0.13
deg$^2$, or 0.2\%, of the CS strip is masked by $R<12$ mag bright stars.  
Thus the total area used for the LF calculation is 64.0 deg$^2$.

\subsection{Magnitude Limits}

	Our samples are magnitude limited to $V_0<16.70$ and $R_0<16.20$
mag.  The median rest-frame color of the $V$ and $R$ samples is
$(\vr)_0=0.53\pm0.07$ mag.  Thus the two samples are of comparable depth.

	We apply galactic extinction and $k$ corrections {\it before}
performing the magnitude selection.  This procedure ensures that we sample
different spectral types to the same depth in each sample.  If we do not
apply $k$ corrections before the magnitude selection, the limiting
absolute magnitude at a particular redshift is a function of spectral
type.  In a purely apparent-magnitude limited redshift survey, late type
galaxies are sampled to fainter absolute magnitudes than early type
galaxies.

	Early redshift surveys, including the APM \citep{loveday92}, CfA
\citep{marzke94}, and SSRS2 \citep{marzke98}, sample mean redshifts
$\overline{z}<0.05$ and have poor photometric precision: $\pm0.3-0.4$ mag.  
$K$ corrections for some of these early surveys are approximated as a
single value for all galaxy types.  Although formally incorrect, the $k$
corrections used for these surveys are much smaller than the photometric
error. More recent surveys, including the LCRS \citep{lin96}, ESP
\citep{zucca97}, and the current 2dF \citep{cross00} and SDSS
\citep{blanton01}, access deeper redshifts $(\overline{z}\simeq0.1$) and
reach photometric accuracies of $\pm0.1-0.2$ mag, yet continue to apply
uncorrected apparent magnitude cuts.  At the 2dF limiting redshift
$z=0.12$ \citep{cross00} there is a $k_B\simeq0.47$ mag spread between
ellipticals and irregulars \citep{frei94}.  As a result, the survey
samples irregulars to a 0.47 fainter absolute magnitude than it does
ellipticals.

	Applying $k$ corrections before the magnitude selection makes very
little difference for our low redshift ($\overline{z}=0.064$) $V$ and $R$
samples.  Although it is impossible to create directly comparable samples,
we calculate LFs for galactic extinction-corrected magnitude limits $m_V =
16.70 + ({\rm median}~k_V)$ and $m_R = 16.20 + ({\rm median}~k_R)$. These
samples contain $<2\%$ fewer galaxies than their pre-$k$ corrected
counterparts, and thus are of similar depth.  The characteristic absolute
magnitudes $M^*$ are $+.01$ to $+.06$ fainter, and the faint end slopes
change by $\pm.01$ (see Equation \ref{schfn} below).  For deeper samples
and bluer passbands the effects are, of course, larger.

\subsubsection{Completeness}

	Completeness, as normally quoted, is the fraction of galaxies with
redshifts in an apparent magnitude-limited sample.  Application of the $k$
correction may bring fainter galaxies without redshifts into the sample.
To minimize this problem, we measured redshifts for galaxies to $R_{\rm
lim} + k_{Rmax}$ and $V_{\rm lim} + k_{Vmax}$.  We have 77\% (303 of 392)
complete redshifts in the interval ($V_{\rm lim}$, $V_{\rm lim} +
k_{Vmax}$), and a similar fraction in $R$. To estimate our true
completeness, we assume that the faint galaxies without redshifts have
approximately the same distribution of $k$ corrections as the sample with
redshifts (Figure \ref{figka}b).  Thus
	\begin{equation} \label{completeness}
	{\rm completeness} = \frac{N_{tot} + 
\int_{m_{\rm lim}}^{m_{\rm lim} + kmax} n(m) f_k(m) dm  }
{ N_{tot} + N_{kmax} },
	\end{equation} where $N_{tot}$ is the total number of galaxies
within our k-corrected magnitude limit $m<m_{\rm lim}$, $n(m)$ is the
number of galaxies per $dm$ fainter than the magnitude limit, $f_k(m)$ is
the fraction of galaxies with $k$ corrections equal to $m$, and $N_{kmax}$
is the total number of galaxies in the interval $(m_{\rm lim}\le m \le
m_{\rm lim} + k_{max})$.

	Equation \ref{completeness} yields 98.1\% completeness in $V$ and
98.2\% completeness in $R$.  The completeness is to the {\it same depth}
for all spectral types.  Furthermore, our long slit spectra have {\it no}
sampling bias in high density regions (e.g.\ because of fiber positioning
constraints) nor is there any bias in surface brightness (as encountered
by some fiber surveys).  The fraction of galaxies without redshifts that
may, given the observed distribution of $k$ corrections, be shifted into
our samples limits our completeness.

\subsection{Redshift Limits}

	We construct LFs for $1,000 \le cz \le 45,000$ km s$^{-1}$, where
$cz$ is the recession velocity in the Local Group frame, $cz = cz_\sun +
300 \sin{l} \cos{b}$, corrected for Virgo-centric infall.  The lower
velocity limit eliminates regions where peculiar velocities may be a large
fraction of the Hubble flow; the upper limit is where the survey becomes
sparse.  Because we calculate recession velocities from the Local Group
frame, we include the effect of Virgo infall. We apply the Virgo-centric
infall model of \citet{kraan86} with the Sandage infall value $v_{vc} =
220$ km s$^{-1}$.  The Virgo infall model only affects the $\sim50$
galaxies in our samples with $cz<6,000$ km s$^{-1}$.  Beyond 6,000 km
s$^{-1}$, the Virgo infall correction is equivalent to an additive dipole
velocity of $\simeq200$ km s$^{-1}$ that makes galaxies appear slightly
brighter.  The Virgo infall correction changes our LFs by making the $V$
and $R$ faint end slopes +0.03 flatter; the characteristic absolute
magnitudes $M^*$ brighten by just -.01 mag (see Equation \ref{schfn} 
below).

\subsection{Luminosities}

	We calculate absolute magnitudes from
	\begin{equation}
M = m - k - A - 5 \log D_L(z) - 25,
	\end{equation} where $m$ is the observed magnitude, $k$ is the $k$
correction, $A$ is the galactic extinction correction, and $D_L$ is the
luminosity distance.  We use the distance measure formulae described by
\citet{hogg99} and compute the luminosity distance for three cosmologies.  
To allow comparisons with earlier work, we use $H_0$=100 km s$^{-1}$
Mpc$^{-1}$, $\Omega_m = 1.0$, and $\Omega_\Lambda = 0.0$ (CDM).  We also
use the flat cosmology $H_0$=75, $\Omega_m = 0.3$, and $\Omega_\Lambda =
0.7$ ($\Lambda$CDM), consistent with recent measurements. To compare with
the original CS open cosmology, we use the open cosmology $H_0$=100,
$\Omega_m = 0.3$, and $\Omega_\Lambda = 0.0$ (OCDM).  Choosing alternative
values for $H_0=100h$ km s$^{-1}$ Mpc$^{-1}$ shifts all absolute
magnitudes by 5log$_{10}h$ and changes number densities by $h^3$.

\section{SAMPLE PROPERTIES}

	Here we examine the luminosities, colors, and surface brightnesses
of the galaxies in our $V$ and $R$ magnitude limited samples.  The data
show strong correlations between luminosity, color, and surface
brightness.

	We examine the population of LSB galaxies in detail.  When we
select LSB galaxies in the same way as \citet{oneil97a}, we find a similar
surface number density of galaxies and distribution of scale lengths and
central surface brightnesses.  This agreement suggests we are complete for
LSB galaxies to our magnitude limit.  When we do not impose a minimum
angular diameter limit (that will select against dwarfs), we find more
blue LSB galaxies than \citet{oneil97a}.  LSB galaxies fall in the large
scale structure marked by high surface brightness galaxies.

\subsection{Color-Surface Brightness Relation}

	Figure \ref{figvru} shows the color-surface brightness relation
for the $R$-selected sample, $(\vr)_0=(-0.11\pm0.05)\mu_{R,0} +
(2.6\pm0.9)$.  The error is the formal error of the least squares fit.  
The rest frame color, $(\vr)_0$, includes galactic extinction and $k$
corrections. The rest frame central surface brightness, $\mu_{R,0}$, is
	\begin{equation}
\mu_{R,0} = \mu_R - 10\log_{10}(1+z) - k_R,
	\end{equation} where $z$ is the redshift and $k_R$ is the $k$
correction.  This relation is based on the sub-sample with SExtractor peak
surface brightnesses, $\mu_R$, avoiding assumptions about fitting light
profile shapes.  We make no corrections for inclination or internal
extinction.  The color-surface brightness relation is valid in the range
$0.35 < (\vr)_0 < 0.60$ mag and $17 < \mu_{R,0} < 21$ mag arcsec$^{-2}$.

	The color-surface brightness relation is not tight.  However, it
disagrees with the complete lack of correlation between color and central
surface brightness claimed by \citet{mcgaugh94} and \citet{oneil97a}. They
measure a continuous range of colors, from very blue to very red, at low
surface brightnesses.  In contrast, Figure \ref{figvru} shows that the
lowest surface brightness galaxies in our sample are among the bluest
galaxies in our sample.  Our deep $R$ images allow a sensitive search for
red LSB galaxies, yet we find very few.

	The disagreement with \citet{mcgaugh94} and \citet{oneil97a} may
be a consequence of their minimum angular diameter.  Angular
diameter-limited surveys are biased toward finding intrinsically large
galaxies; only very nearby dwarfs are included in such surveys. From deep
CCD images of the Cancer and Pegasus clusters and fields in the Great Wall
region, \citet{oneil97a} cataloged 127 LSB galaxies with $\theta >
26\arcsec$.  We do not impose a minimum angular diameter. We thus include
more low surface brightness blue dwarfs.

\subsection{Color-Luminosity Relation}

	Figure \ref{figvrm} shows the color-absolute magnitude relation
for the $R$-selected galaxies, $(\vr)_0 = (-0.08\pm0.03) M_R -
(1.13\pm0.5)$.  The error is the formal error of the least squares fit.
$M_R$ is the $R$ band absolute magnitude. The scatter about the
color-absolute magnitude relation is large in part because we make no
corrections for inclination or internal extinction.

	\citet{visvanathan81} measures color-absolute magnitude relations
for 55 Virgo and field spirals in the $V$, $r$, and $IV$ filters.  His
passbands allow a rough comparison with our data, though we cannot convert
the narrow (300\AA\ at 6738 \AA) $r$ filter to Kron-Cousins $R$ to obtain
a proper zero-point.  There are numerous measurements of the elliptical
galaxy color-magnitude relation, but these are not comparable with our
field galaxy sample.  \citet{visvanathan81} finds a $(V-r)$ slope of
$-0.06\pm0.04$, similar to our result.

\subsection{Luminosity-Surface Brightness Relation}

	Figure \ref{figum} shows the luminosity-surface brightness
relation for the $R$-selected galaxies, $M_R = (2.1\pm0.7)\mu_{R, 0}
-(60.3\pm2)$.  

	For comparison, the 2dF survey finds an identically strong
luminosity-surface brightness relation in the $b_J$ band, with slope
$2.4_{-0.5}^{+1.5}$ \citep{cross00}.  \citet{driver99} uses a
volume-limited sample of 47 galaxies from the Hubble Deep Field and finds
$M_{F450W}=(1.5\pm0.2)\mu_e-(50\pm2)$. Here $\mu_e$ is the rest frame,
mean surface brightness within the half light (``effective'') radius of
the 25 $I$(F814W) mag arcsec$^{-2}$ isophote.  Figure \ref{figum} shows
the \citet{driver99} luminosity-surface brightness relation assuming
$(B-R)=1.2$ \citep{fukugita95}.  The \citet{driver99} luminosity-surface
brightness slope is shallower than ours, but we are unable to discriminate
between the two relations.  \citet{cross00} do not publish a zero-point,
preventing a comparison here.


\subsection{Low Surface Brightness Galaxies}

	We explore the surface brightness completeness of our sample by
comparing it with the \citet{oneil97a} 26$\arcsec$ angular
diameter-limited LSB galaxy catalog.  \citet{oneil97b} publish $B$
photometry and $(\bv)$ colors.  We thus use the $V$ catalog for the
comparison.

	We follow \citet{oneil97a} and determine the central surface
brightness $\mu(0)_V$ and scale length $\alpha_V$ of LSB galaxies by
fitting exponential functions to the $V$ surface brightness profiles.  To
compare with \citet{oneil97a} we assume $\alpha_B=\alpha_V$. Our
experience shows this is assumption is acceptable; $V$ scale lengths
are $\sim5\%$ larger than our $R$ scale lengths.  We note that if
$\alpha_B>\alpha_V$, the 26$\arcsec$ diameter limit will include fewer
galaxies in the $V$ sample than in the $B$ sample.  We use the exponential
profile to find the radius of the LSB galaxies at the limiting sky
brightness $\mu_{sky}=26$ mag arcsec$^{-2}$ with $r =
\alpha_V/1.086~(\mu_{sky} - \mu(0)_V).$ The comparison with
\citet{oneil97a} also assumes $\mu(0)_V = \mu(0)_B - (\bv)$.  This is a
rough approximation because we assume the average (\bv)=0.79 mag for the
40\% of the \citet{oneil97b} sample without (\bv) colors.
 
	Figure \ref{figlsb} shows $\mu(0)_V$ vs.\ $\alpha_V$ for LSB
galaxies which satisfy $V < 16.70$ mag, $\mu(0)_V>21.2$ mag arcsec$^{-2}$,
and $\theta > 26\arcsec$ in the \citet{oneil97a} catalog and in
the $V$ sample. The solid and dashed lines show the limiting magnitude
$V_{\rm lim}=16.70$ for a face-on, pure exponential disk galaxy for
limiting sky brightnesses 26 and 28 mag arcsec$^{-2}$.  Galaxies
satisfying the magnitude limit should fall above the lines; galaxies which
fall below are LSB galaxies with central peaks or extended wings poorly
fit by an exponential profile.  \citet{oneil97a} lack redshifts.  Their
central surface brightnesses and magnitudes, although galactic extinction
corrected, are not rest-frame corrected.  The effect should be small
because the redshifts should be small.

	Figure \ref{figlsb} shows that our LSB galaxies cover the same
range of scale length and central surface brightness as the
\citet{oneil97a} galaxies.  A Kolmogorov-Smirnov test gives a 50\%
probability of the surface brightnesses being drawn from the same sample,
and a 11\% probability of the scale lengths being drawn from the same
sample.  A significant fraction of the scale lengths in our sample
are smaller than those in \citet{oneil97a}.  Applying the $26\arcsec$
diameter limit at the sky limit may not be a good comparison. If we
instead apply the $26\arcsec$ diameter limit at $\mu_{sky}=25$ mag
arcsec$^{-2}$, our smallest scale length is approximately equal to the
smallest \citet{oneil97a} scale length and a K-S test gives a 67\%
probability that the scale lengths are drawn from the same sample.

	Figure \ref{figlsb} indicates that bright magnitude-limited
surveys using CCD photometry are unlikely to miss LSB galaxies.  LSB
galaxies with central surface brightness near the limiting sky brightness
must also have extreme scale lengths in order to be bright enough for
inclusion in bright magnitude-limited samples.  Such galaxies are rare in
Figure \ref{figlsb}. The mean scale length of LSB galaxies in the
\citet{oneil97a} sample is $8\arcsec$.  A galaxy with an $8\arcsec$
exponential scale length has the same $V=16.7$ isophotal magnitude with a
central surface brightness only 0.20 mag arcsec$^{-2}$ fainter for
limiting sky brightnesses of 26 and 28 mag arcsec$^{-2}$ (see Figure
\ref{figlsb}).  This inclusion of essentially all appropriate LSB galaxies
in our sample is in agreement with the \cite{deJong2000} local Sb-Sdm
galaxy bivariate distribution functions.

	Although we cannot compare the space density of LSB galaxies with
\citet{oneil97a}, we can compare the surface number density to a given
limiting magnitude.  For $V<16.70$, $\mu(0)_V > 21.2$, and $\theta >
26\arcsec$, the \citet{oneil97a} catalog contains 44 LSB galaxies over 27
deg$^{2}$ or 1.6 deg$^{-2}$; we have 40 LSB galaxies over 64 deg$^{2}$ or
0.63 deg$^{-2}$. \citet{oneil97a} find a factor of 2.6 more LSB galaxies
per deg$^{2}$ than we do.

	Clustering in the galaxy distribution probably explains the
apparent deficiency of LSB galaxies in the CS.  \citet{oneil97a} search
for LSB galaxies in the Pegasus and Cancer clusters and around known
galaxies in the Great Wall region.  The average galaxy density in our
sample is much lower than in these regions.  If LSB galaxies are clustered
like high surface brightness galaxies, as \citet{mo94} claim for the CfA
and $IRAS$ surveys, we can account for the difference. We contrast counts
of galaxies in 0.56 deg$^2$ fields \citep[the typical field size used
by][]{oneil97a} centered on known CS galaxies with counts in fields
positioned at random.  The count around known galaxies in the CS catalog
is 2.3 times the count in the random fields.  Thus accounting for
clustering brings the surface number density of LSB galaxies in our sample
in agreement to $\sim15\%$ with the \citet{oneil97a} measurements.

	Figure \ref{figzoom} shows the $V$-limited redshift survey to
12000 km s$^{-1}$; {\it all} LSB galaxies fall in the large-scale
structure marked by the high surface brightness galaxies.  \citet{obs00}
draw a similar conclusion for galaxies in the Pegasus and Cancer clusters.

	The LSB galaxies in the CS are mostly at low redshift and thus are
intrinsically faint.  Figure \ref{figmr} shows the absolute magnitudes of
the $R$ sample as a function of redshift; circles mark the LSB galaxies.
The intrinsic faintness of the LSB galaxies is consistent with the
absolute magnitude-surface brightness relation.  The 1000 km/s velocity
limit eliminates a third of the LSB galaxies (Figure \ref{figzoom}) from
the LF calculation.

\section{LUMINOSITY FUNCTIONS}

	We compute the LF using two related methods unbiased by density
inhomogeneities in the galaxy distribution:  the parametric
maximum-likelihood method of \citet[][hereafter STY]{sty} and the
nonparametric stepwise maximum-likelihood method of \citet[][hereafter
EEP]{eep}.  Others have described the details of the calculations
\citep[e.g.][]{lin96}.  We do not reproduce them here.  We assume that the
LF is independent of position, and determine its shape and amplitude
separately.

	We assume a \citet{schechter76} parameterization of the LF,
	\begin{equation} \label{schfn} \phi(M) = 0.4 \ln(10) \phi^* 
~[10^{0.4(M^*-M)}]^{(1+\alpha)} ~\exp(-10^{0.4(M^*-M)}),
	\end{equation} where $\phi^*$ is the normalization (galaxies per
unit volume), $\alpha$ is the faint end slope, and $M^*$ is the
characteristic absolute magnitude.  However, the observed LF is a
convolution of the true LF and the photometric error.  In the STY
calculations, we make the approximation that our errors are Gaussian
distributed and use the Schechter LF convolved with the total photometric
error.  The error is a sum in quadrature of the zero-point,
measurement, galactic extinction, and $k$ correction errors.

\subsection{Results}

	Figure \ref{figbigslice} shows the full $R$ magnitude-limited
redshift slice.  Figures \ref{figcsr} and \ref{figcsv} show the STY
(curve) and EEP (symbols with error bars) estimates of the LF for the $R$
and $V$ magnitude limited samples in the OCDM cosmology.  The
inset boxes shows the $2\sigma ~\chi^2$ error ellipses for $M^*$ and
$\alpha$. Table \ref{cs} summarizes the results.  Schechter function
parameters for the OCDM cosmology are:  $M^*_R=-20.88\pm0.09$,
$\alpha_R=-1.07\pm0.09$, and $\phi^*_R=0.016\pm0.003$ Mpc$^{-3}$ and
$M^*_V=-20.23\pm0.09$, $\alpha_V=-1.07\pm0.09$, and
$\phi^*_V=0.020\pm0.003$ Mpc$^{-3}$.

	Surprisingly, the $V$ and $R$ local galaxy LF faint end slopes are
identical.  These LFs apply to $V$ and $R$ magnitude-limited samples to
comparable depth of the same region of sky. From the color-magnitude
relation (Figure \ref{figvrm}), we expect that the $V$ sample would lose
red, intrinsically bright galaxies and gain blue, intrinsically faint
galaxies, thus producing a steeper faint end slope.

	To examine the behavior of these samples, Figures \ref{figcsbr}
and \ref{figcsbr2} show the LFs of the bluest 1/3 and reddest 1/3 of the
$R$ and $V$-selected samples.  We divide the sample in thirds to separate
the blue and red colors cleanly, given the typical $\pm0.07$ mag (\vr)
error.  Figure \ref{figvrhist} plots the distribution of rest frame
$(\vr)_0$ colors, with lines marking the $R$ $(\vr)_0<0.499$ and
$(\vr)_0>0.555$ mag and $V$ $(\vr)_0<0.494$ and $(\vr)_0>0.551$ mag color
cuts.  The reddest third generally includes galaxy types Sab and earlier;
the bluest third generally includes galaxy types Sc and later.

	The bright end of the LF is primarily composed of red, luminous,
early-type galaxies and the faint end is composed of blue, intrinsically
faint, late-type galaxies.  The reddest third LFs show an excess of
extremely luminous $M_V < -22$ cD galaxies and contain no galaxies less
luminous than $M_V > -18$.  The bluest third of the $V$ and $R$-selected
samples show no excess of luminous galaxies, but instead determine the
entire faint end slope.  For $M_V>-17.5$, 50\% of the galaxies in the
magnitude-limited samples are LSB galaxies.

	Our $R$ LF agrees with the original CS LF to within the $1\sigma$
error of the original CS.  The appropriate comparison, however, is with
the CS-West LF \citep{geller97}; our 64 deg$^2$ covers the west half of
the CS and does {\it not} include the Corona Borealis region.  Compared to
the CS-West LF, the new $R$ faint end slope $\alpha_R$ is +0.10 shallower,
the new $M^*_R$ is -0.15 mag brighter, and the new amplitude $\phi^*_R$ is
35\% lower.  The differences, although not formally significant, are
attributable to the cleaner sample and refinements of the analysis.

\subsection{Discussion}

\subsubsection{The $V$ and $R$ faint end slopes}

	Here we attempt to determine why the $V$ and $R$ faint end slopes
appear to be independent of passband.  We examine the galaxies unique to
the $V$ and $R$ samples and study their effect on the LF determinations.

	44 of 1255 $V$-selected galaxies are unique to the $V$ sample
(triangles in Figures \ref{figvru} and \ref{figvrm}), and 40 of 1251
$R$-selected galaxies are unique to the $R$ sample (squares in Figures
\ref{figvru} and \ref{figvrm}).  The 44 $V$ galaxies have lower peak
surface brightness and luminosity than the 40 $R$ galaxies.  The mean
luminosity of the 44 $V$ galaxies is $\overline{M_V}=-18.9\pm1.4$; the
mean luminosity of the 40 $R$ galaxies is $\overline{M_R}=-20.8\pm0.8$.  
Consistent with the color-magnitude relation, the 44 $V$ galaxies are
significantly bluer than the 40 $R$ galaxies.  Mean rest-frame colors are
$\overline{(\vr)}_0=0.39\pm0.06$ mag for the 44 $V$ galaxies and
$\overline{(\vr)}_0=0.59\pm0.04$ mag for the 40 $R$ galaxies.  Figures
\ref{figcsbr} and \ref{figcsbr2} show that the faint end slope of the $V$
bluest third is -0.14 steeper than the $R$ bluest third.  Similarly, the
$V$ reddest third faint end slope is shallower than the $R$ reddest third.

	The $V$ and $R$ faint end slopes probably appear identical because
the weight of the LF fit comes from the knee of the LF around $M^*$, and
$M^*$ and $\alpha$ are strongly correlated.  $\alpha$ and $M^*$ have a
correlation coefficient of $\rho_{\alpha,M^*}\simeq0.9$
\citep{schechter76}.  Figure \ref{figcs2} illustrates that the $V$- and
$R$-only galaxies contribute to the LF largely around $M^*$, but represent
only 3\% of the galaxies within 1 magnitude of $M^*$.  We conclude the
that identical $V$ and $R$ faint end slopes result from fitting the 97\%
$M^*$ galaxies common to both samples.

	This result suggests that a magnitude-limited LF is a rather blunt
tool for characterizing the faint end of the local galaxy population.  
Furthermore, the faint end comes from a necessarily small volume in a
magnitude-limited survey.  The population of low luminosity galaxies in
the sample is sensitive to local large scale structure. A volume-limited
sample of galaxies, complete to low luminosities, would provide a much
better determination of the faint end slope of the galaxy LF.

	Surveys of clusters, including Virgo \citep{sandage85} and Coma
\citep{bernstein95}, probe the galaxy LF to faint luminosities.  Although
it has become clear in recent years that dwarfs contribute to steep faint
end slopes, there is little consensus on the slope.  Published values of
the faint end slope in clusters and groups range from $-1.2$ to less than
$-2.0$, and various authors \citep[e.g.][]{driver96,trentham98} make a
convincing case for a flat bright end slope and an upturn around $-18 <
M_B < -16$ mag.  In these studies, cluster membership is based primarily
on statistical background subtraction for lack of redshifts.  A faint end
slope as steep as $-2$ was recently supported by \citet{kambas00} who
assume every LSB galaxy in their images is a member of the Fornax cluster.  
The numerical simulations of \citet{valotto01} show there is a strong
tendency for projection effects to cause an apparent rising faint end
slope in clusters.  Although our samples are limited to $M_V < -17$ mag,
we note that our results are consistent with a faint end slope as steep as
$\alpha \simeq-1.2$.

\subsubsection{Luminosity density}

	A fundamental quantity provided by the galaxy LF is the local
luminosity density of the Universe,
	\begin{equation}
j = \phi^* L^* \Gamma(\alpha+2).
	\end{equation} We use $M_{V \sun}=+4.82$ and $M_{R \sun}=+4.28$
mag \citep{allen00}.  Table \ref{cs} summarizes the local luminosity
densities. For the OCDM cosmology, $j_R=1.9\pm0.6\times10^8 h L_\sun$
Mpc$^{-3}$ and $j_V=2.2\pm0.7\times10^8 h L_\sun$ Mpc$^{-3}$.

	In the luminosity range $-23 < M_V < -17$ mag, LSB galaxies appear
to contribute little to the total luminosity density of the local
universe.  The magnitude-surface brightness relation shows that the bulk
of LSB galaxies are of intrinsic low luminosity, and our shallow faint end
slope limits the number of LSB galaxies that might contribute to the
luminosity density.  Our samples provide no constraints on whether LSB
galaxies make a larger contribution to the local luminosity density at
fainter absolute magnitudes.

	Figure \ref{figj} compares the luminosity densities and limiting
magnitudes of various surveys (Table \ref{lfs}).  The surveys use a common
CDM cosmology; choosing alternative values for $H_0=100h$ km s$^{-1}$
Mpc$^{-1}$ shifts $j$ by the factor $h$. We compare luminosity densities
in the $b_J$ band.  We convert our $V$ sample to $b_J$ using the typical
Sab color $(b_J-R)=1.2$ of \citet{fukugita95}, plus our mean
$(\vr)_0=0.53$.  We convert the LCRS with $(b_J-R)=1.1$ \citep{lin96}.  
The CfA2 Zwicky magnitudes are adjusted by $(b_J-m_Z)=-0.35$
\citep{gaztanaga00}.  We note that no correction to the CfA amplitude is
needed.  Both \citet{bothun90} and \citet{grogin99} find no significant
photometric scale error in samples of 107 and 230 northern Zwicky
galaxies.


	Figure \ref{figj} shows the effect of large scale structure on
measurements of the local luminosity density.  The northern surveys, which
sample the adjacent Virgo cluster and Great Wall region, measure a
systematically higher luminosity density than southern surveys of
comparable depth.  This discrepancy is a result of the absence of nearby
rich clusters in the south \citep[i.e.][]{marzke98}.  The ESP and 2dF
appear to go deep enough to sample beyond this southern under-dense
region.

	The CS luminosity density is comparable to the luminosity density
measured by the northern redshift surveys and by the deepest southern
redshift surveys.  The CS luminosity density is identical with the
northern CfA and SDSS surveys when variation in luminosity density over
large scale is accounted for.  The rich Corona Borealis region in the
eastern half of the original CS contains 75\% more luminosity density than
the western half plotted in Figure \ref{figj}.  When averaged together,
the $b_J$ luminosity density of the full CS is $\simeq3.0\pm0.7\times10^8
h L_\sun$ Mpc$^{-3}$, nearly identical to the CfA and SDSS results.

\subsubsection{Comparison with other luminosity functions}

	In Figure \ref{figall} we compare the CS $R$-band local galaxy LF
with other CCD based surveys and with the 2dF survey.  The LCRS has Gunn
$r$ photometry calibrated in Cousins $R$; we offset $M^*$ by $(R-r)=-0.1$
mag \citep{shectman96}.  To compare the SDSS $r^*$-band LF, we shift $M^*$
by $(R-r^*)=-0.22$ \citep{fukugita96}.  The 2dF is a $b_J$
photographic-based survey; we shift $M^*$ by $(b_J-R)=1.2$ mag.  1.2 mag
is the typical $(b_J-R)$ of an Sab galaxy \citep{fukugita95} and is
consistent with the mean color of the SSRS2 \citep{marzke97}.  We plot the
2dF with its $b_J$ normalization, though this is probably too large for
the $R$ band.  We find a 30\% increase in normalization $\phi^*$ from our
$R$ to $V$ samples, similar to the 30\% increase in normalization from
$r^*$ to $g^*$ in the SDSS \citep{blanton01}. All four LFs in Figure
\ref{figall} share a common CDM cosmology.

	The CS LF parameters $M^*$, $\alpha$, and $\phi^*$ agree at the
$1\sigma$ level with the 2dF and SDSS LF parameters.  On the other hand,
the CS disagrees with the LCRS $M^*$ at the $5\sigma$ level and $\alpha$
at the $4\sigma$ level.  The probable cause of this disagreement is the
surface-brightness limit the LCRS impose on their survey.  The LCRS
probably samples only the bright end of the luminosity-surface brightness
relation, leading to a deficiency of galaxies in the faint end.  
Furthermore, the LCRS uses isophotal photometry which may lead to an
underestimate of the normalization \citep{blanton01}.  The CS, 2dF, and
SDSS make corrections to total magnitudes and have statistically common
results.  We note that the 2dF and SDSS have systematically steeper faint
end slopes and larger normalizations than the CS which may result from
the greater depth of these surveys, or from the k-correction bias (see
section 3.1) which tends to add galaxies to the faint end of the LF.

	The faint end slopes of our color selected sub-samples are
consistent with those found by \citet{marzke97}. \citet{marzke97} measure
plate-based $(B-R)$ colors for a complete, $B_{26}<14.5$ sub-sample of the
SSRS2. Making a cut at $(B_{26}-R_{B26})=1.3$, approximately the color of
an Sbc galaxy, they find faint end slopes of $-1.46\pm0.18$ and
$-0.72\pm0.24$ for the blue and red halves of their sample.  We find
$-1.29$ to $-1.43\pm0.15$ faint end slopes for the bluest third of our
samples, and $-0.21$ to $-0.13\pm0.25$ faint end slopes for the reddest
thirds.  The reddest thirds are closer to the $-0.26$ faint end slope of
reddest quartile of \citet{marzke97}.  The LFs for the red and blue
sub-samples are consistent with the overall picture laid out by
\citet{binggeli88}, who describe the galaxy populations which contribute
the total galaxy LF.

\section{CONCLUSIONS}

	We construct clean $1255$ and $1251$ member $V_0<16.7$ and
$R_0<16.2$ magnitude-limited samples of galaxies from 64 deg$^2$ of
photometry of the Century Survey \citep{geller97}. CCD photometry allows
improvements in resolution, linearity, and sensitivity to LSB galaxies;
average photometric errors are $\pm0.049$ mag.  Colors allow more accurate
$k$ corrections.  We apply $k$ corrections before the magnitude selection
is performed to ensure all spectral types are sampled to the same depth.  
The redshifts, obtained from long-slit spectra, are complete even in high
density regions and have no surface brightness constraints (as encountered
by some fiber surveys). The samples cover $8\fh5 < \alpha_{\rm B1950} <
13\fh5$, $29^\circ < \delta_{\rm B1950}< 30^\circ$ in the northern sky,
and are 98\% complete.

	We examine the properties of the galaxies in our $V$ and $R$
samples, emphasizing the characteristics of the LSB population. We observe
strong correlations between color, surface brightness, and luminosity, and
construct the following relations for the $R$ sample:
$(\vr)_0=(-0.11\pm0.05)\mu_{R,0} + (2.6\pm0.9)$, $(\vr)_0 =
(-0.08\pm0.03) M_R - (1.13\pm0.5)$, and $M_R = (2.1\pm0.7)\mu_{R,0}
-(60.3\pm2)$.

	The LSB galaxy surface number density and distribution of scale
lengths and central surface brightnesses are consistent with the angular
diameter-limited LSB catalog of \citet{oneil97a}, suggesting we are
complete for LSB galaxies to our magnitude limit.  The LSB galaxies all
fall in large scale structure marked by high surface brightness galaxies.

	The faint end slopes of the $V$ and $R$ local galaxy LFs are
surprisingly identical ($\alpha=-1.07$, OCDM), probably because of the
97\% $M^*$ galaxies common to the $V$ and $R$ samples and because of the
strong correlation between $\alpha$ and $M^*$ that dominates the LF
determination.  The OCDM local luminosity densities are:  $j_R=1.9\pm0.6
\times 10^8 h L_\sun$ and $j_V=2.2\pm0.7 \times 10^8 h L_\sun$.  The
luminosity densities are dependent on large scale structure, and are
consistent with the northern (CfA and SDSS) redshift surveys and the
deepest southern (2dF and ESP) redshift surveys. In the luminosity range
$-23 < M_V < -17$ mag, LSB galaxies appear to contribute little to the
total luminosity density of the local universe.

\acknowledgements

	We thank Scott Kenyon, Ken Rines, Perry Berlind, and Mike Calkins
for their assistance with the observations, and Susan Tokarz for
processing the FAST spectra.  We also thank Gary Wegner, John Thorstensen,
and Ronald Marzke for their careful work creating the Century Survey. This
research was supported by the Smithsonian Institution.

\bibliographystyle{apj}
\bibliography{wbrown}

\clearpage
\centerline{\includegraphics*[scale=0.5]{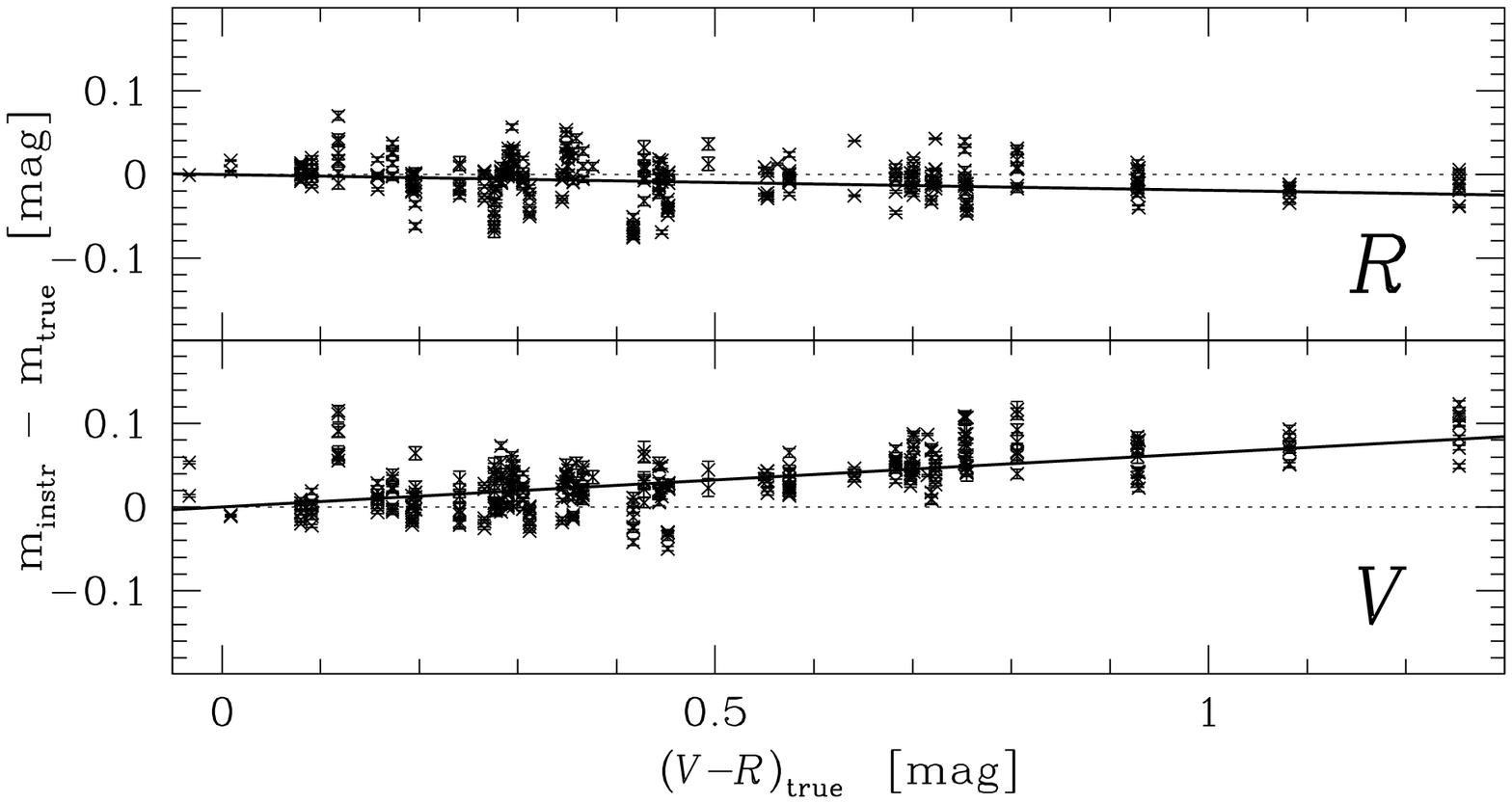}}
\figcaption[wbrown.fig1.ps]{ \label{figct} MOSAIC chip
2 $R$ and $V$ color terms, including data from all nights.}

\centerline{\includegraphics*[scale=0.5]{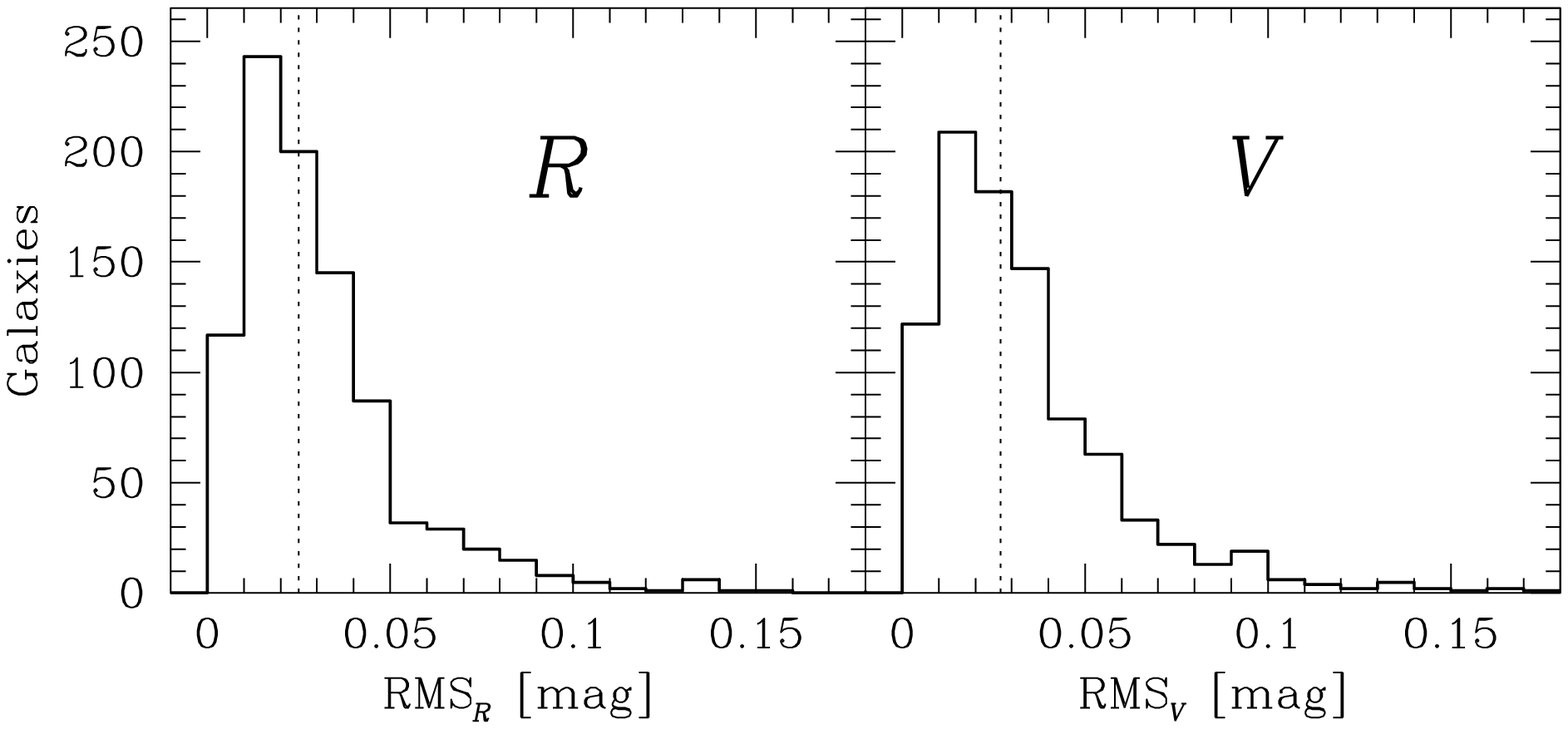}}
\figcaption[wbrown.fig2.ps]{ \label{figerr}
Distribution of RMS photometric errors, for galaxies with $\ge3$
observations.  Dotted lines show the median values RMS$_R=\pm0.025$ and
RMS$_V=\pm0.027$ mag.}

\centerline{\includegraphics*[scale=0.5]{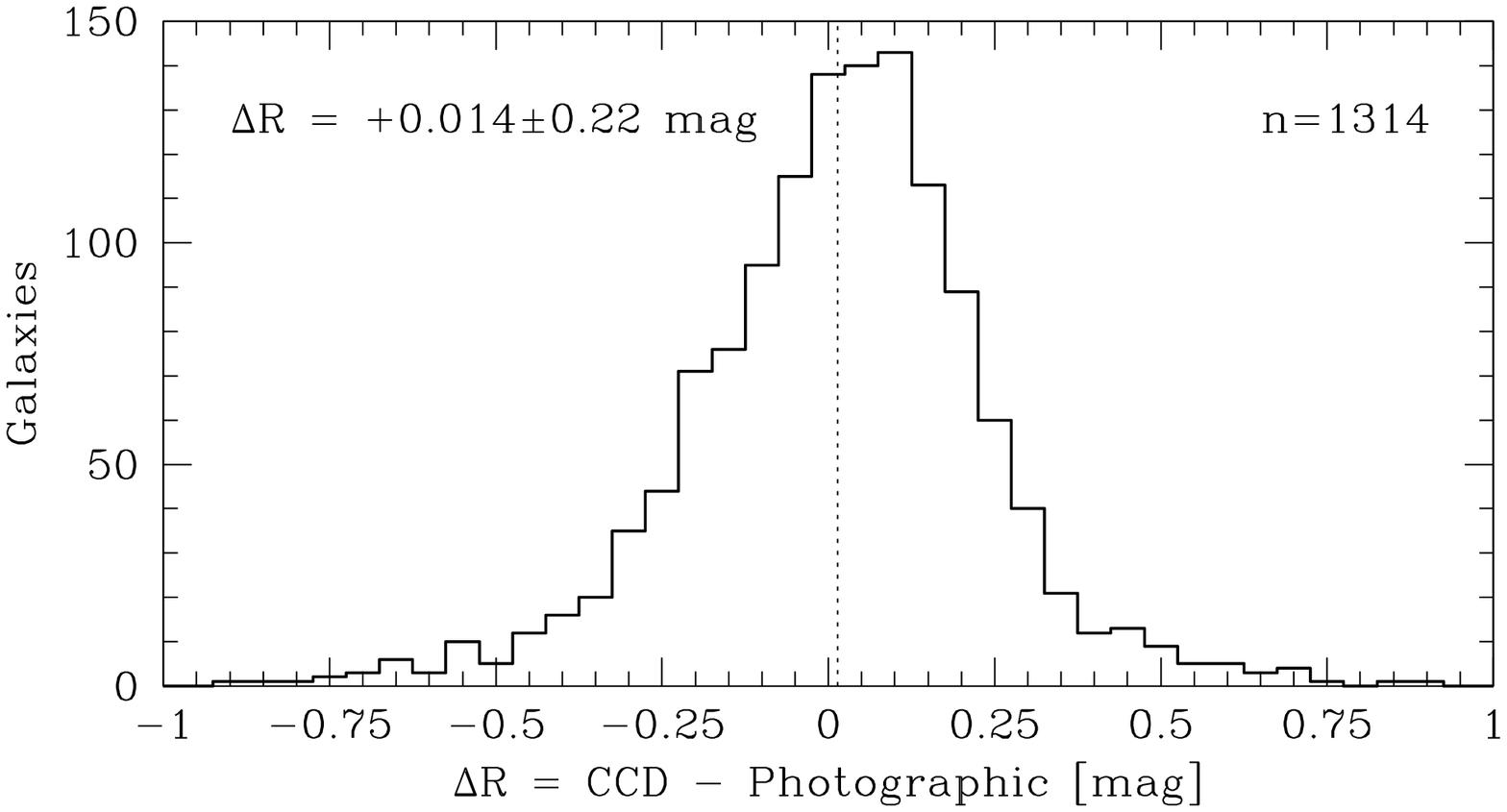}}
\figcaption[wbrown.fig3.ps]{ \label{figcscomp}
Comparison of Century Survey CCD and photographic photometry.}

\centerline{\includegraphics*[scale=0.5]{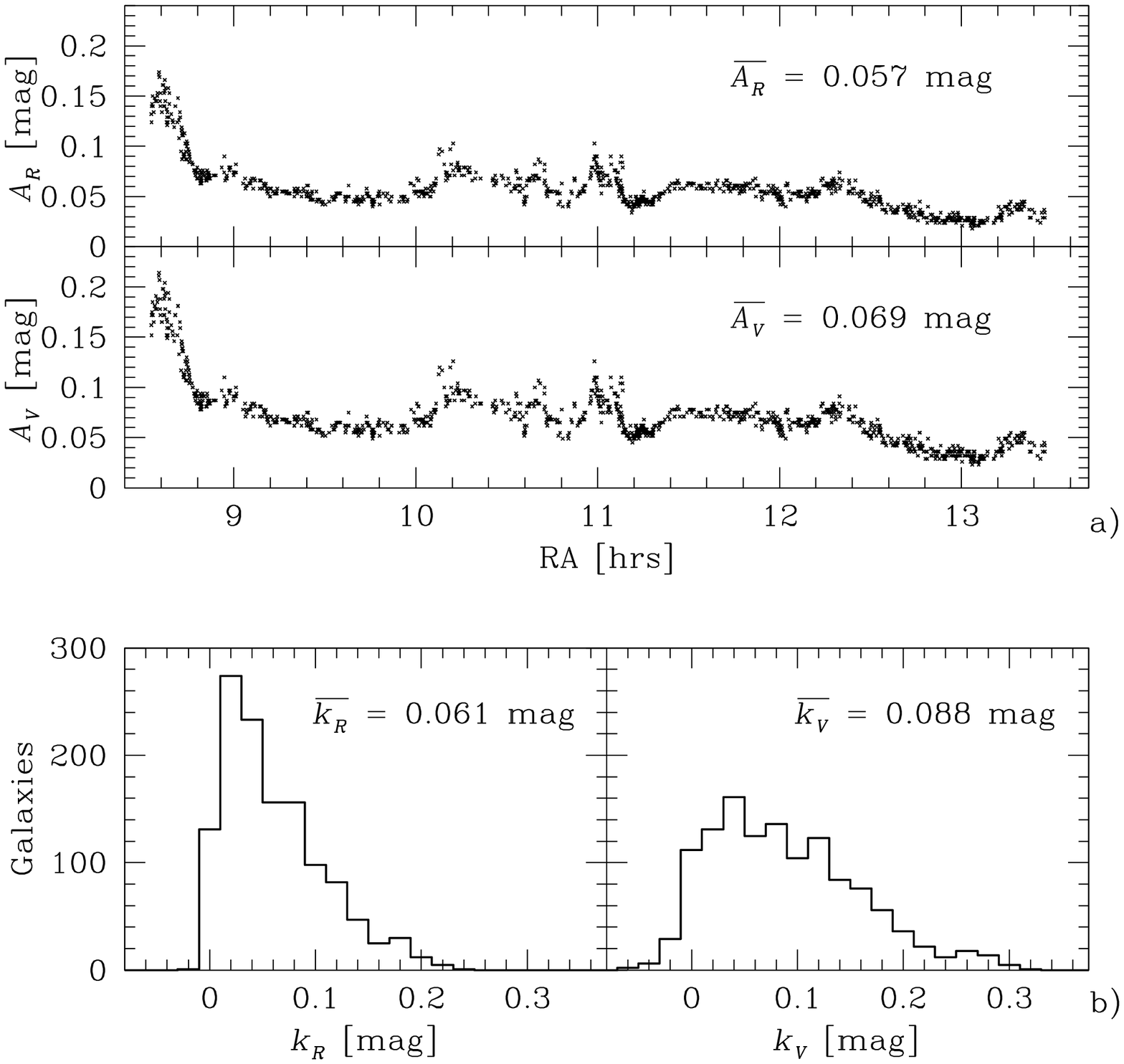}}
\figcaption[wbrown.fig4.ps]{ \label{figka}
a)  Galactic extinction and b) $k$ corrections.}

\centerline{\includegraphics*[scale=0.5]{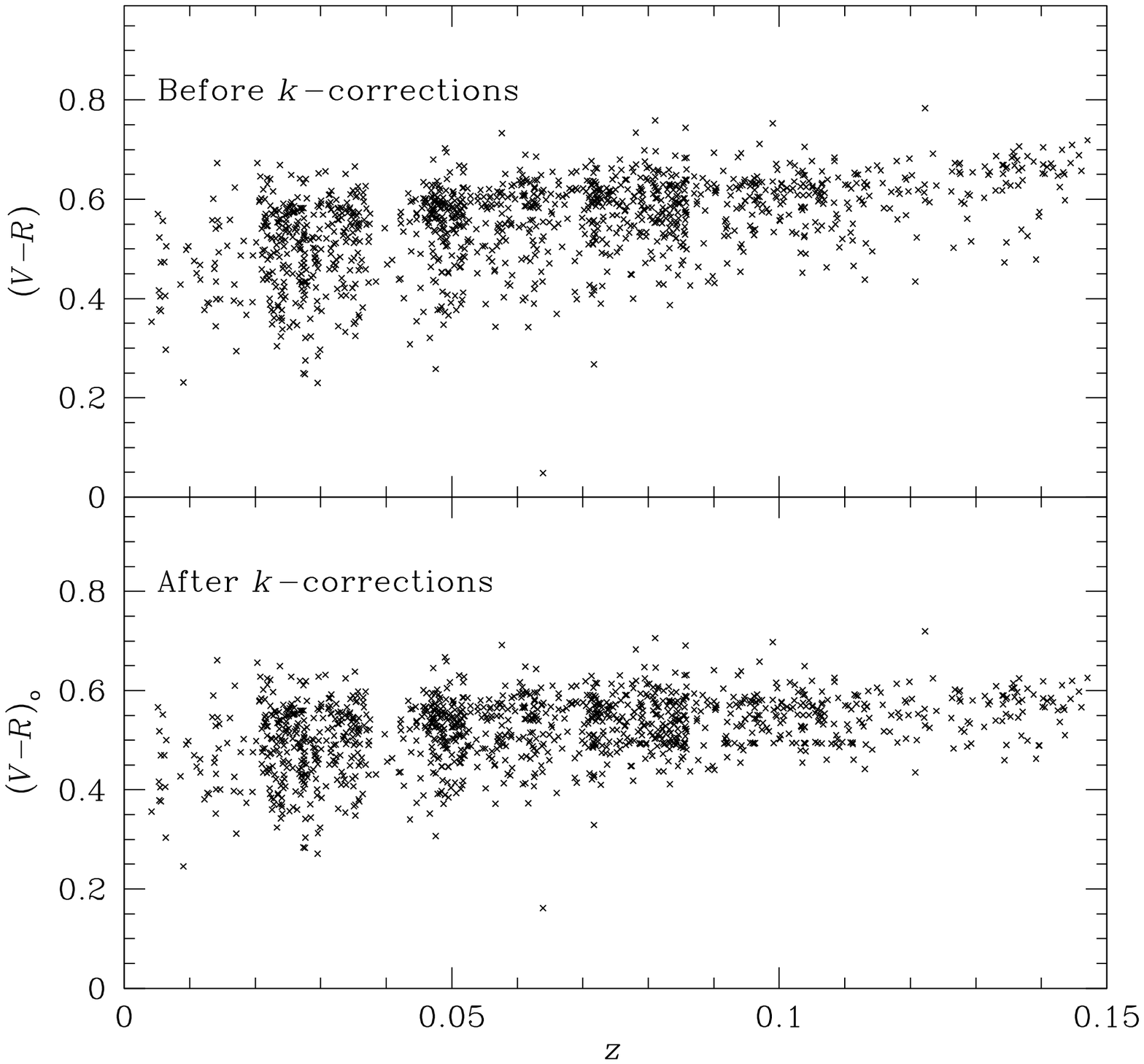}}
\figcaption[wbrown.fig5.ps]{ \label{figkvr}
Distribution of (\vr) colors before and after $k$ corrections for the
$R$-selected sample.}

\centerline{\includegraphics*[scale=0.5]{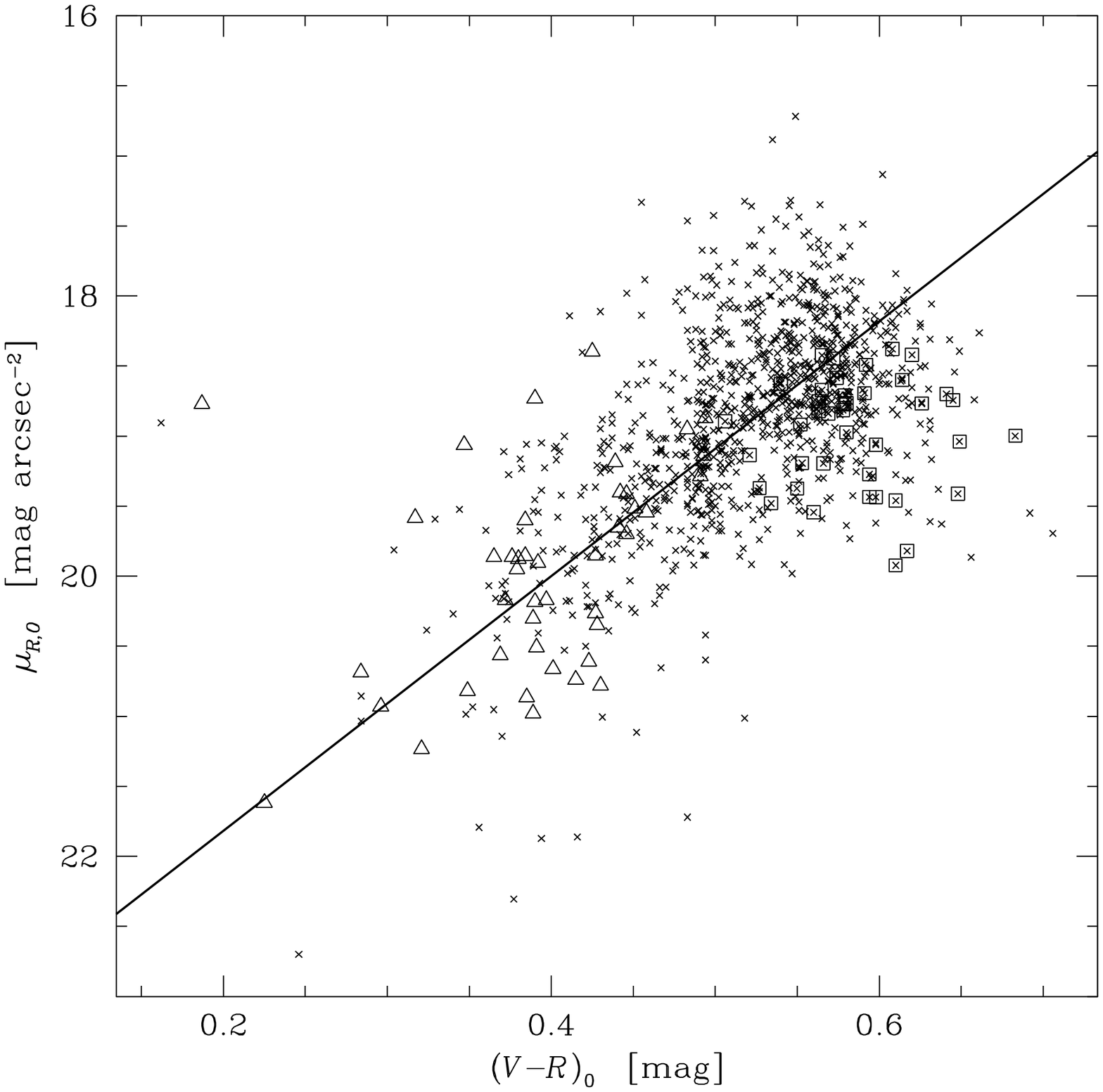}}
\figcaption[wbrown.fig6.ps]{ \label{figvru} Rest
frame color-peak surface brightness relation for the $R$-selected sample:
$(\vr)_0=(-0.11\pm0.05)\mu_{R,0} + (2.6\pm0.9)$.  Open squares mark the 40
galaxies unique to the $R$ sample, open triangle mark the 44 galaxies
unique to the $V$ sample.}

\centerline{\includegraphics*[scale=0.5]{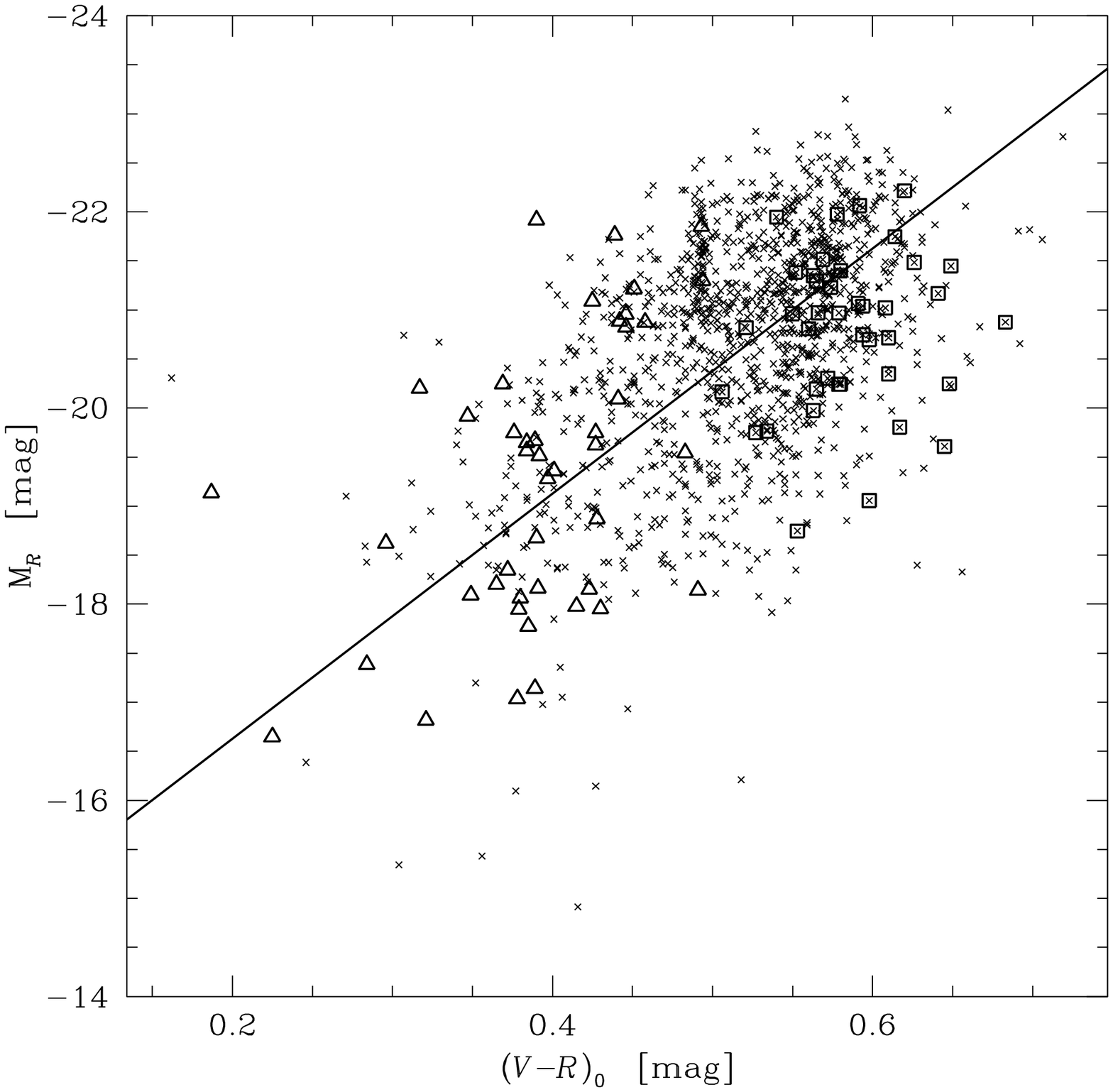}}
\figcaption[wbrown.fig7.ps]{ \label{figvrm} Rest
frame color-absolute magnitude relation for the $R$-selected sample:
$(\vr)_0 = (-0.08\pm0.03) M_R - (1.13\pm0.5)$.  Open squares mark the 40
galaxies unique to the $R$ sample, open triangle mark the 44 galaxies
unique to the $V$ sample.}

\centerline{\includegraphics*[scale=0.5]{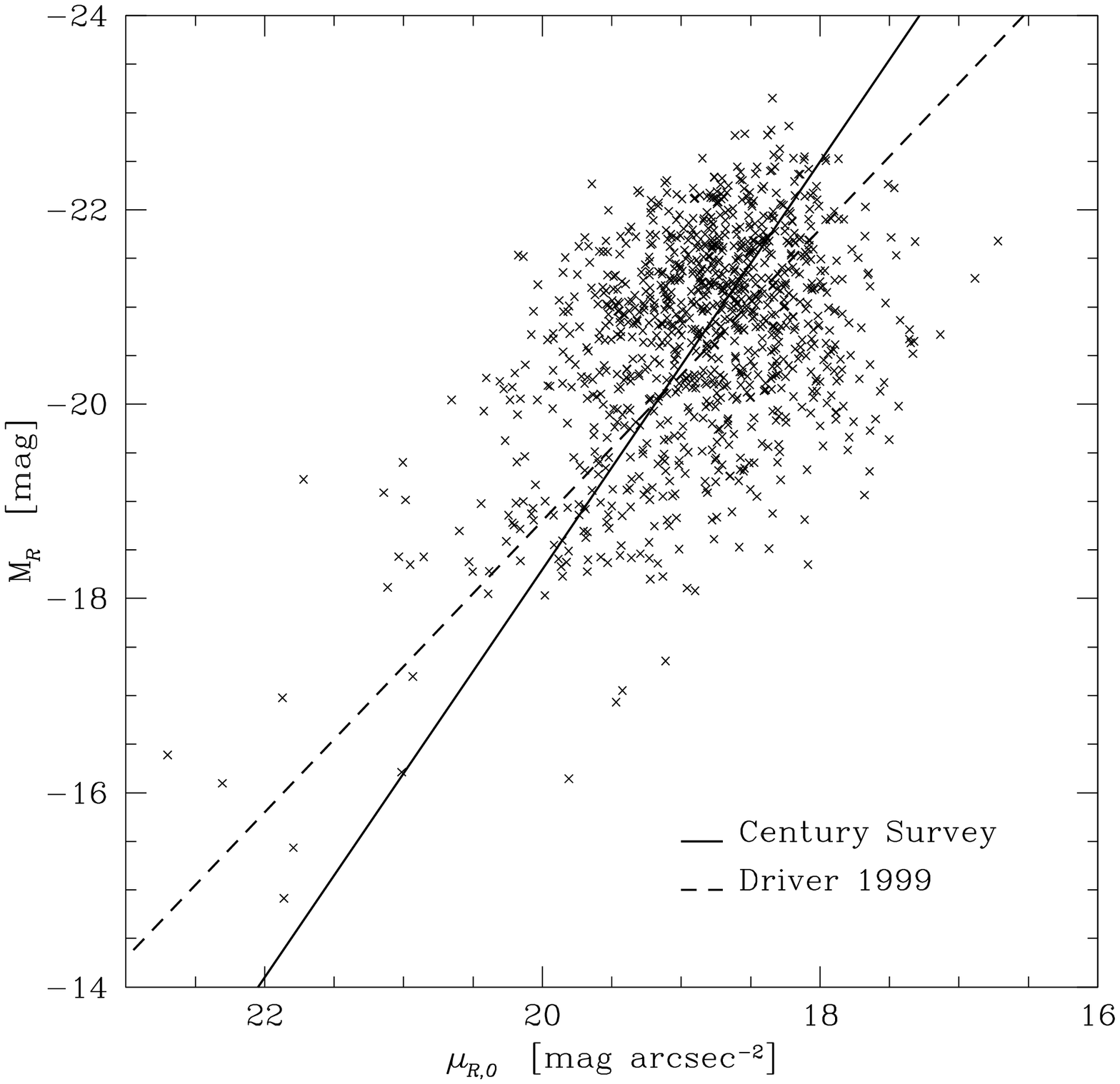}}
\figcaption[wbrown.fig8.ps]{ \label{figum} Absolute
magnitude-surface brightness relation for the $R$-selected sample: $M_R
= (2.5\pm0.7)\mu_{R,0} -(68\pm2)$.}

\centerline{\includegraphics*[scale=0.5]{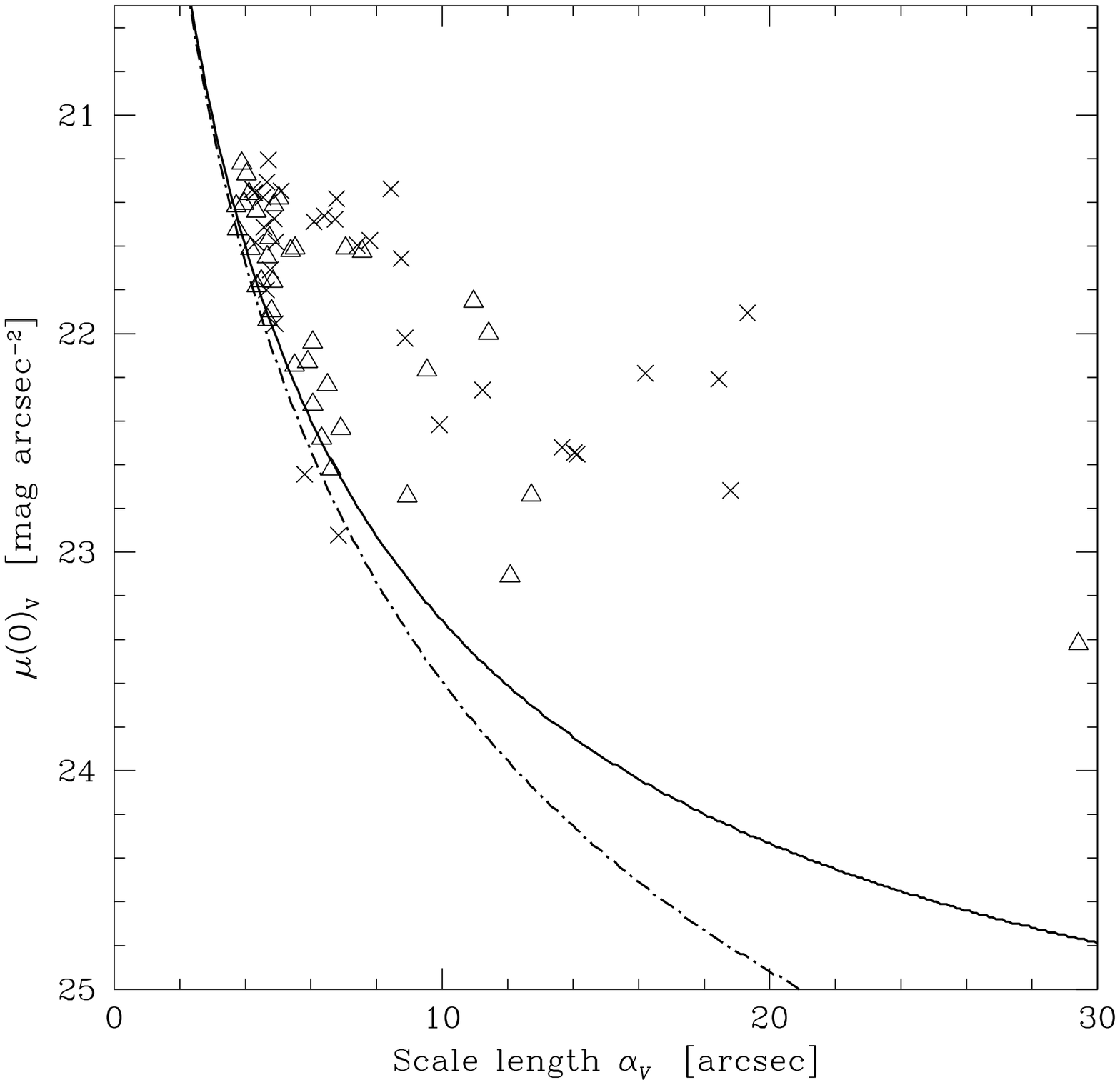}}
\figcaption[wbrown.fig9.ps]{ \label{figlsb} Surface
brightness vs.\ scale length comparison of galaxies satisfying $V<16.70$
mag, $\mu(0)_V>21.2$ mag arcsec$^{-2}$, and $\theta >26\arcsec$ for the
\citet{oneil97b} LSB catalog (x's) and our $V$-selected sample
(triangles).  The lines show the $V=16.70$ limiting magnitude for a
face-on, pure exponential disk profile with limiting sky surface
brightnesses of 26 (solid) and 28 (dash-dot) mag arcsec$^{-2}$.}

\centerline{\includegraphics*[scale=0.5]{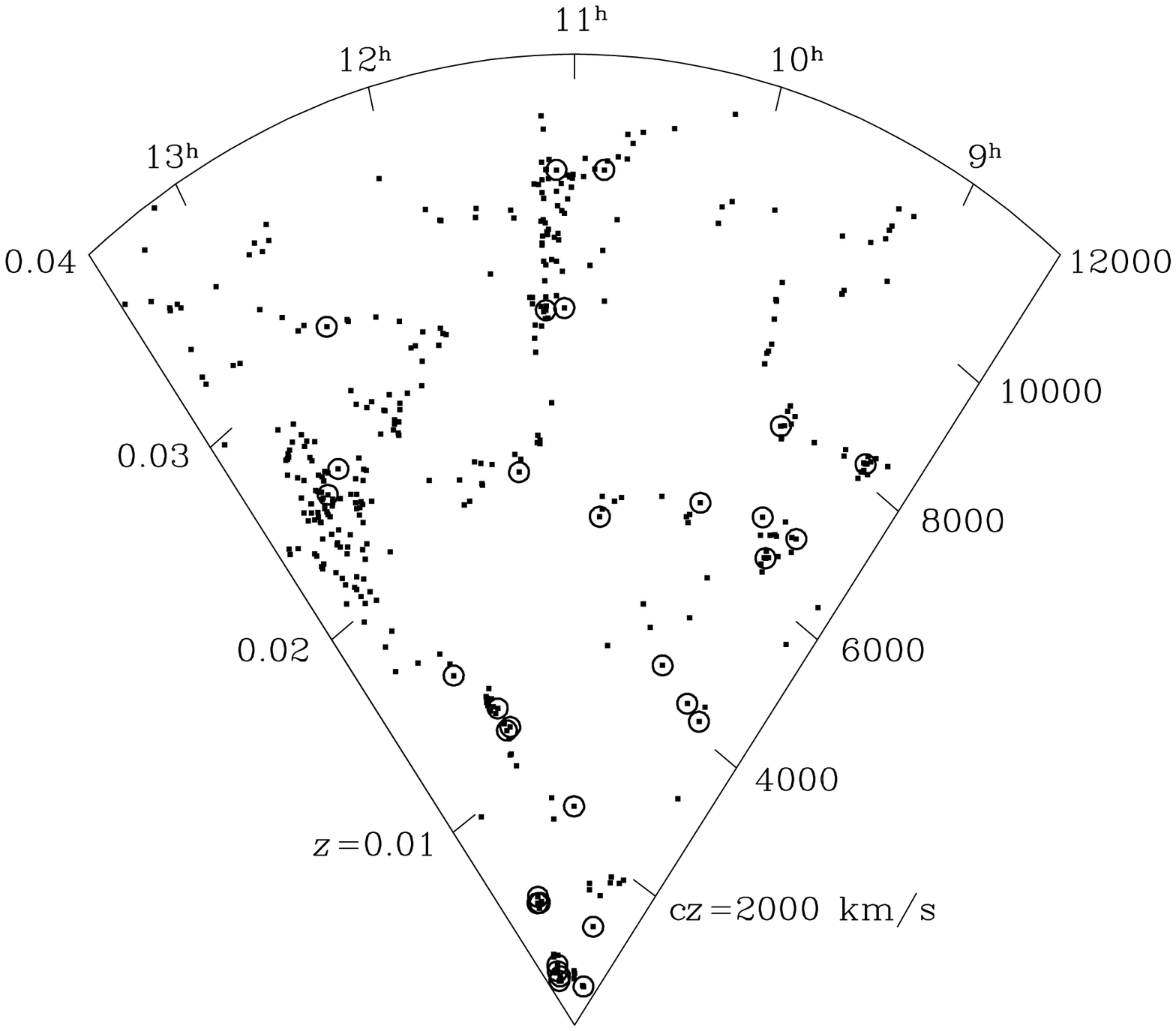}}
\figcaption[wbrown.fig10.ps]{ \label{figzoom}
Cone diagram for the $V_0<16.70$ Century Survey sample to $z=0.04$.  All
low surface brightness galaxies $\mu_V(0)_0 > 21.2$, marked with circles,
fall in large scale structure marked by high surface brightness galaxies.}

\centerline{\includegraphics*[scale=0.5]{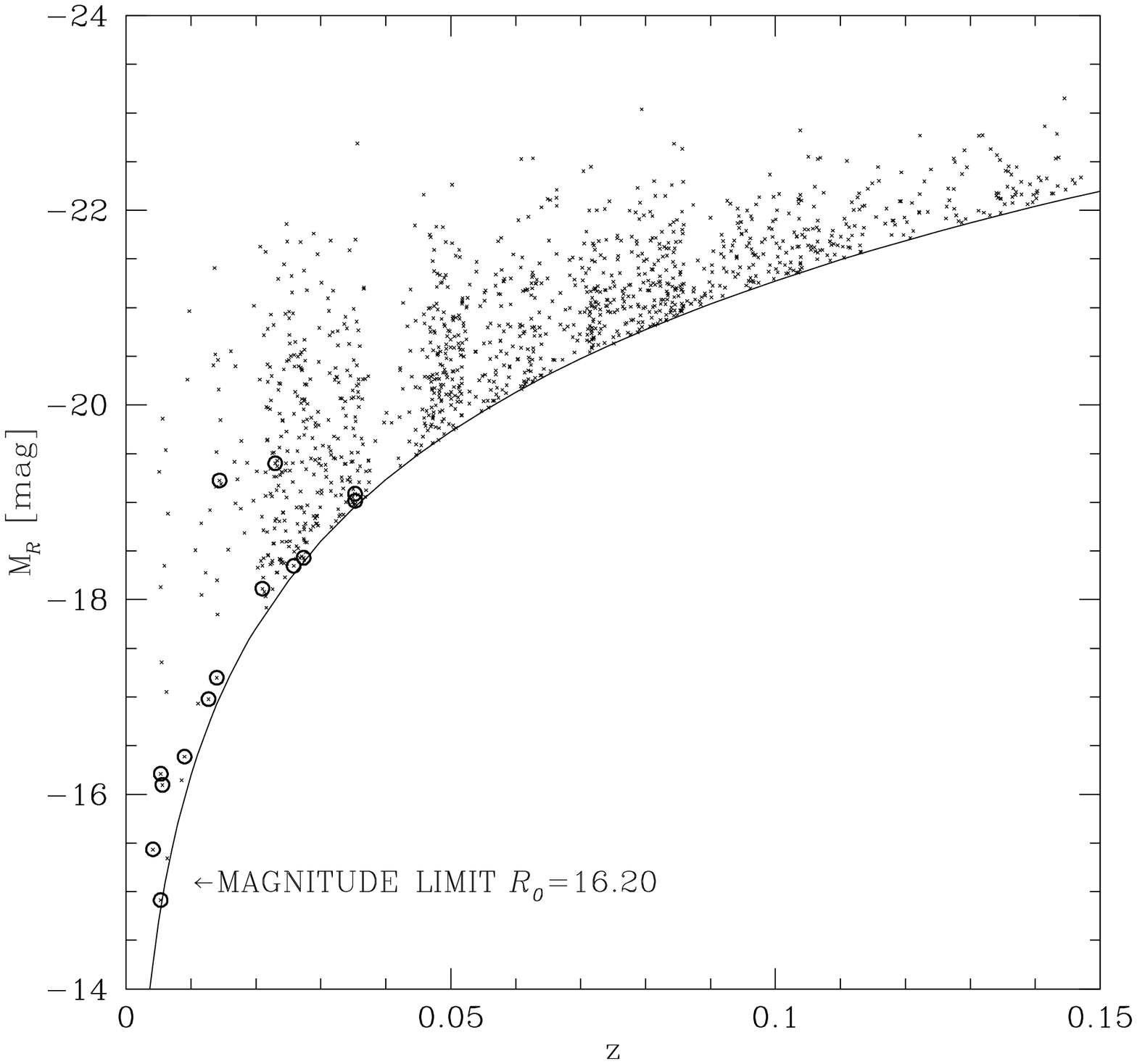}}
\figcaption[wbrown.fig11.ps]{ \label{figmr}
Distribution of absolute magnitude with redshift.  Low surface brightness
galaxies $\mu_{R, 0} > 20.8$, marked with circles, are clearly low
luminosity systems.  The line shows the limiting $R$ absolute
magnitude vs.\ redshift.}

\centerline{\includegraphics*[scale=0.5]{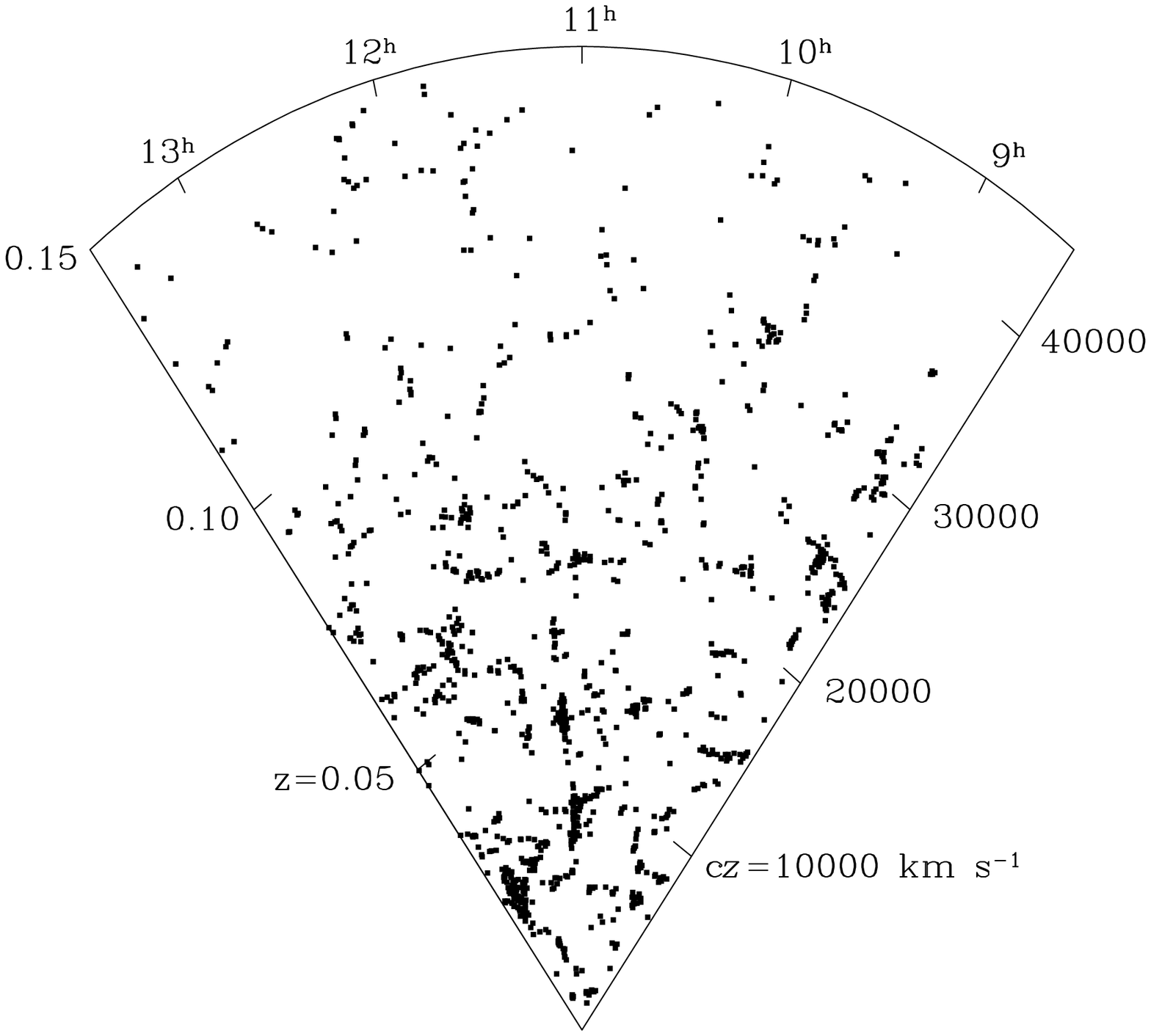}}
\figcaption[wbrown.fig12.ps]{ \label{figbigslice}
Cone diagram for the full $R_0<16.20$ Century Survey sample.}

\centerline{\includegraphics*[scale=0.5]{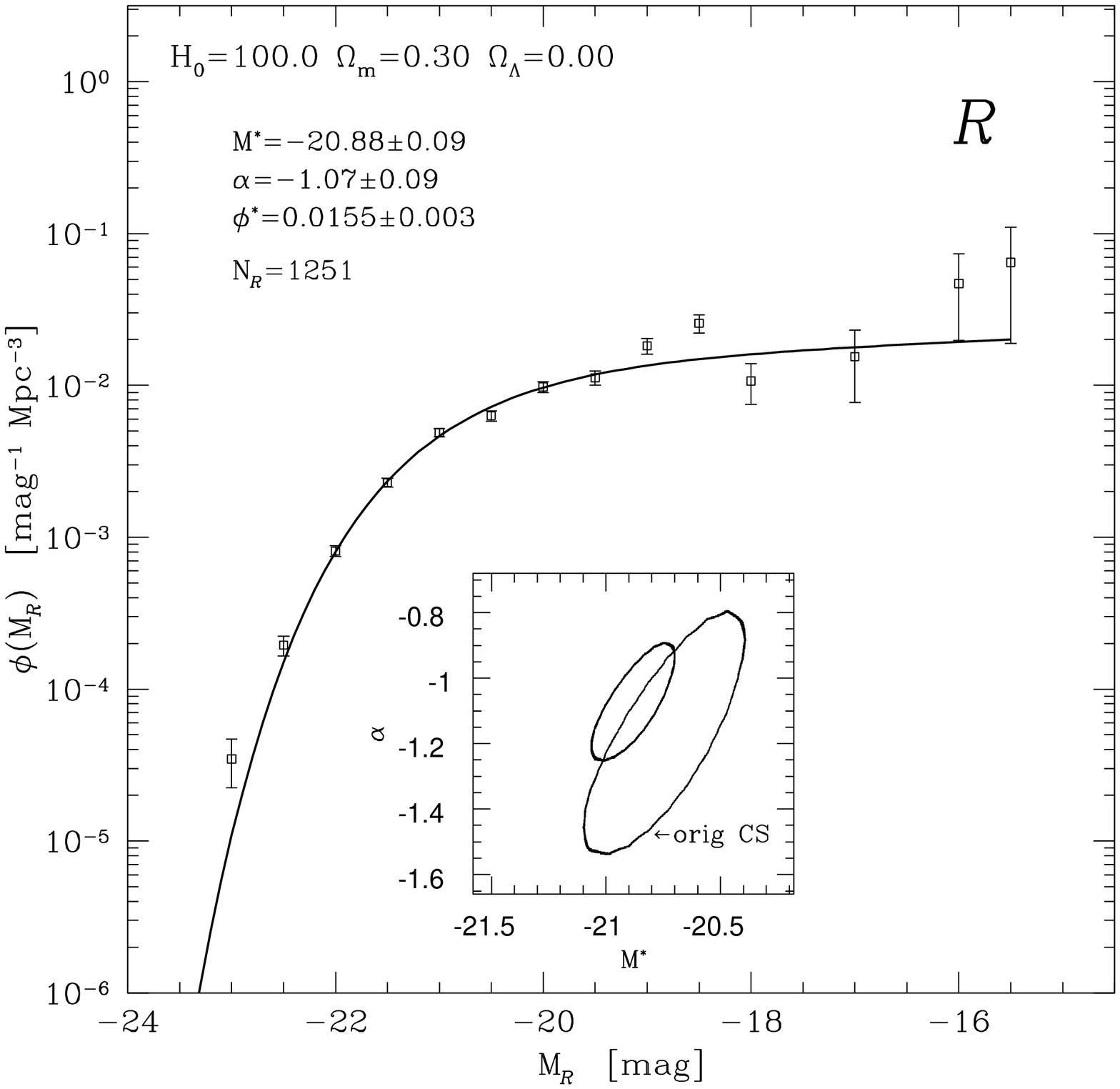}}
\figcaption[wbrown.fig13.ps]{ \label{figcsr}
$R_0 < 16.20$ galaxy luminosity function (OCDM).  Inset shows the $2
\sigma ~\chi^2$ error ellipse for $M^*$ and $\alpha$.}

\centerline{\includegraphics*[scale=0.5]{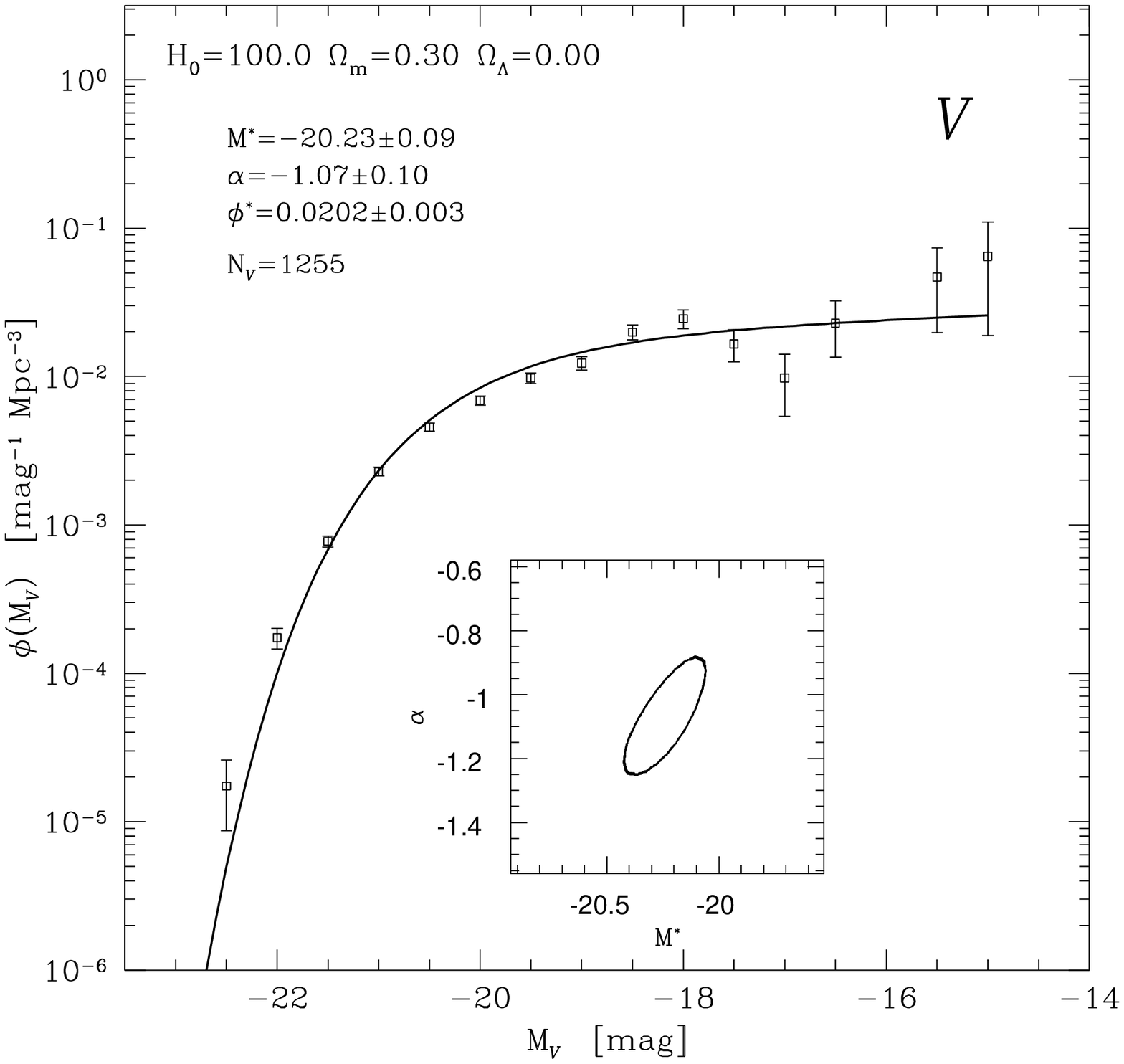}}
\figcaption[wbrown.fig14.ps]{ \label{figcsv} $V_0 <
16.70$ galaxy luminosity function (OCDM). Inset shows the $2\sigma
~\chi^2$ error ellipse for $M^*$ and $\alpha$.}

\centerline{\includegraphics*[scale=0.5]{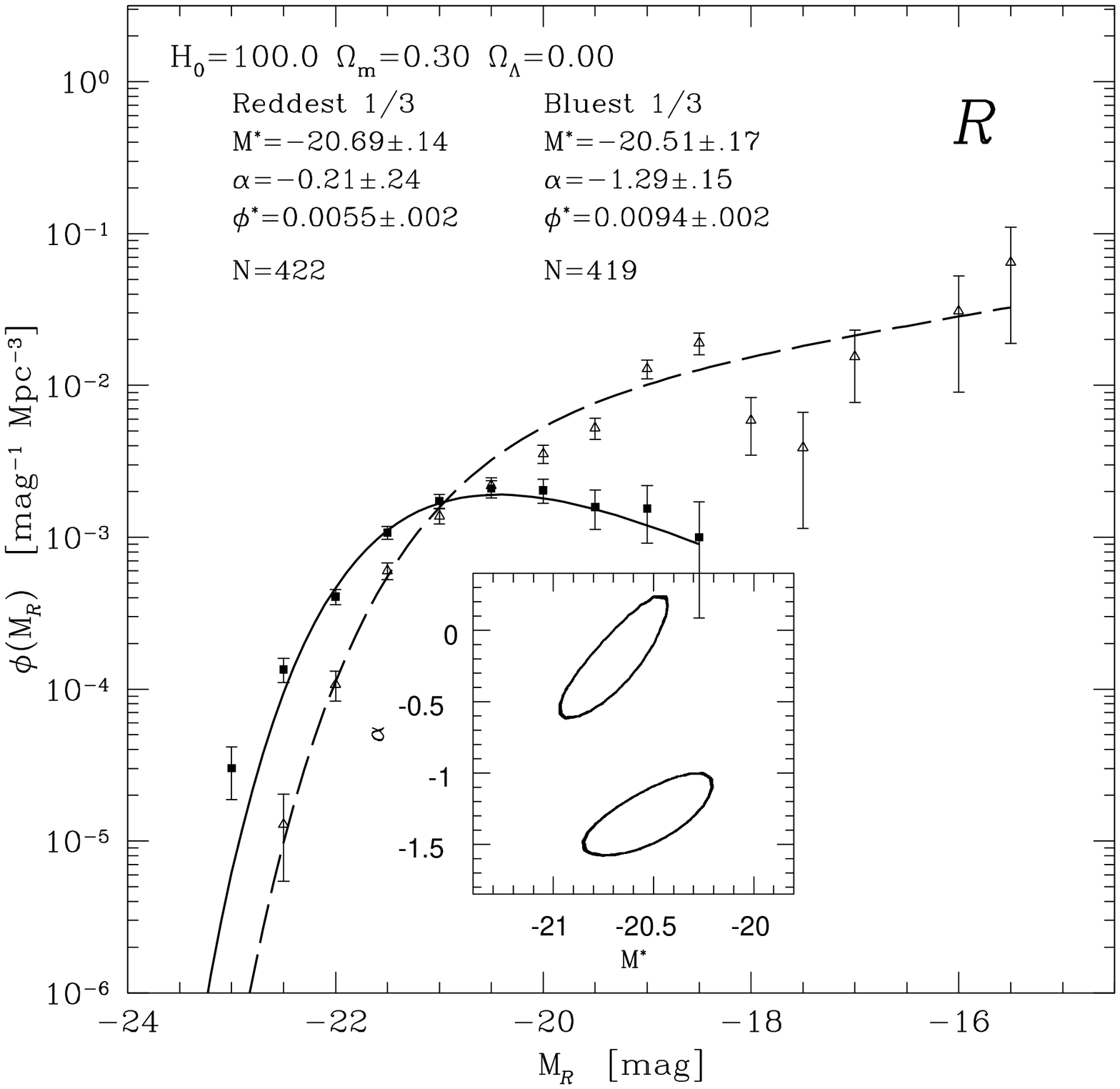}}
\figcaption[wbrown.fig15.ps]{ \label{figcsbr}
Luminosity functions of the reddest 1/3 (solid line) and bluest 1/3
(dashed line) of the $R$-selected sample (OCDM).  Inset shows the $2\sigma
~\chi^2$ error ellipses for $M^*$ and $\alpha$.}

\centerline{\includegraphics*[scale=0.5]{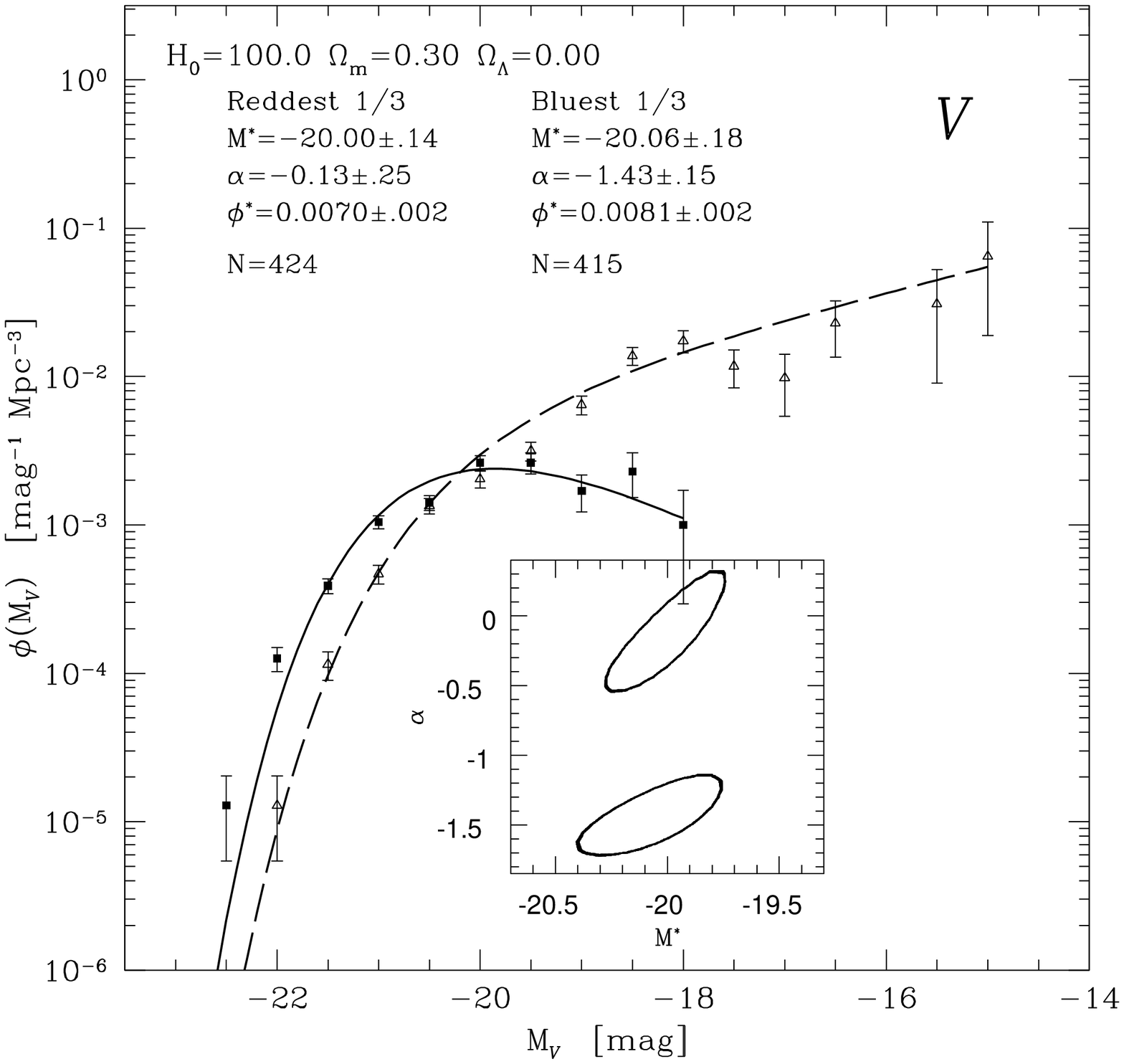}}
\figcaption[wbrown.fig16.ps]{ \label{figcsbr2}
Luminosity functions of the reddest 1/3 (solid line) and bluest 1/3
(dashed line) of the $V$-selected sample (OCDM).  Inset shows the $2\sigma
~\chi^2$ error ellipses for $M^*$ and $\alpha$.} 

\centerline{\includegraphics*[scale=0.5]{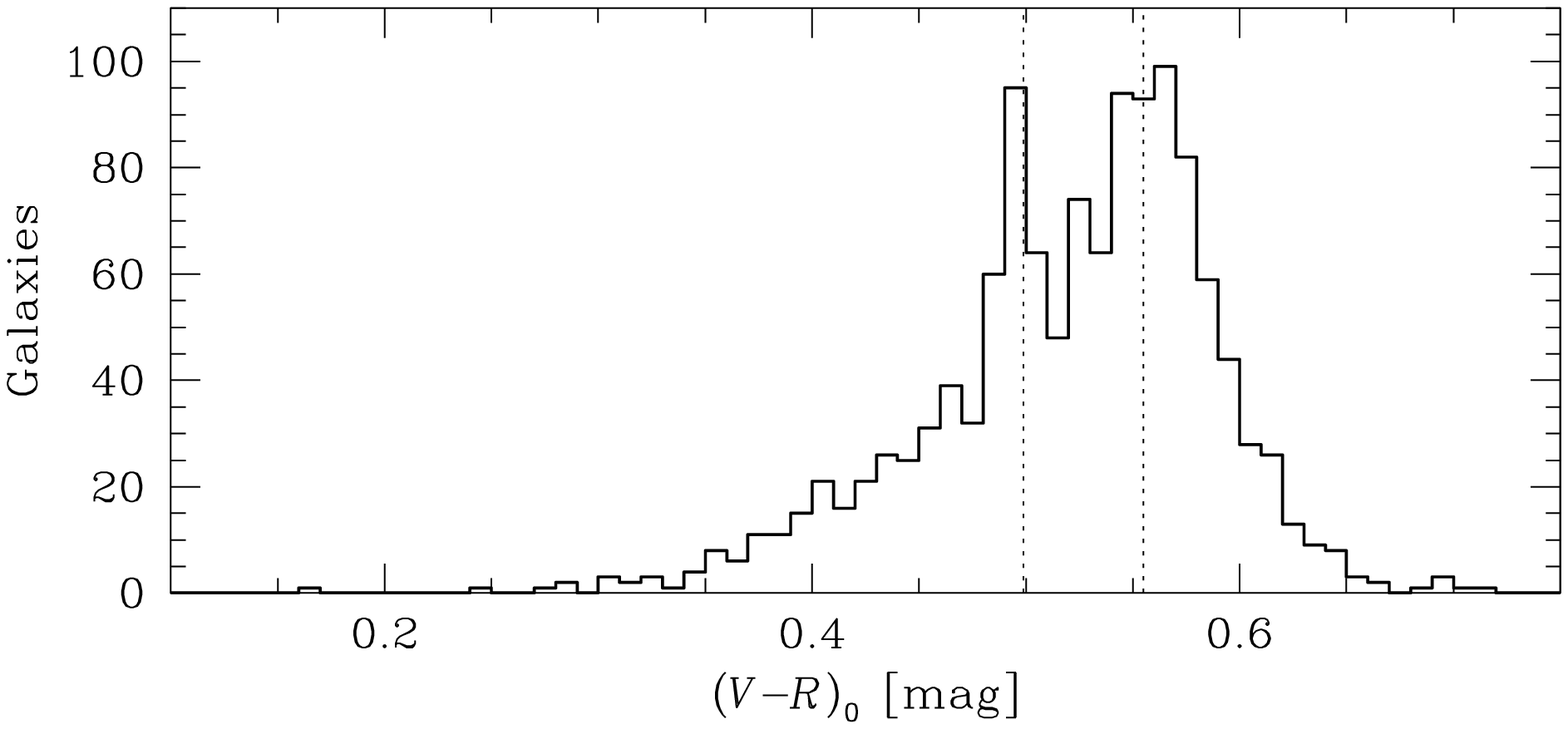}}
\figcaption[wbrown.fig17.ps]{ \label{figvrhist}
Histogram of the rest-frame $(\vr)_0$ colors.  Dotted lines mark the
bluest and reddest thirds of the $R$-selected sample.}

\centerline{\includegraphics*[scale=0.5]{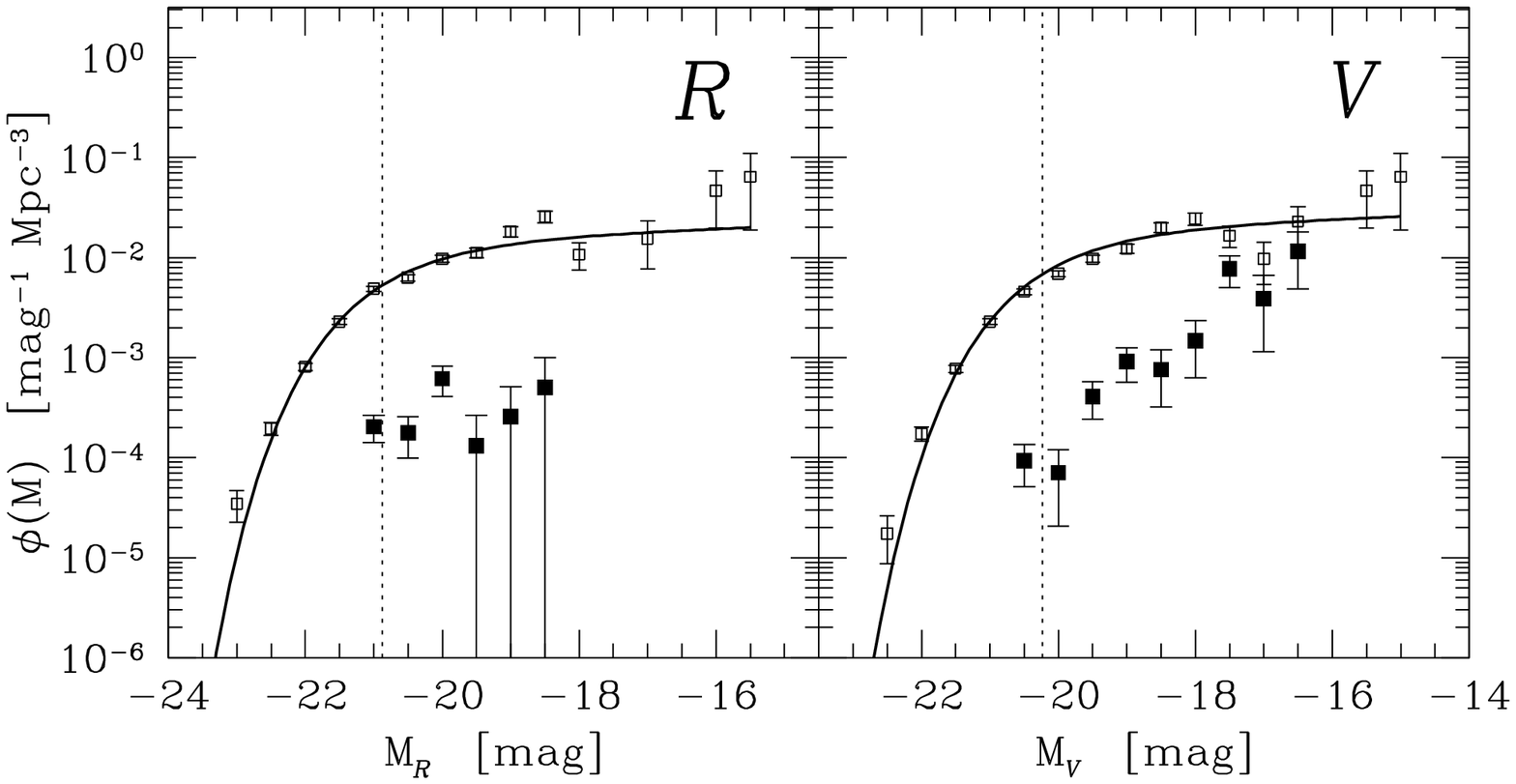}}
\figcaption[wbrown.fig18.ps]{ \label{figcs2}
Luminosity distribution of the galaxies unique to the $V$ and $R$ samples
(solid squares) compared with the full samples.  The dotted line marks
$M^*$ (OCDM).}

\centerline{\includegraphics*[scale=0.5]{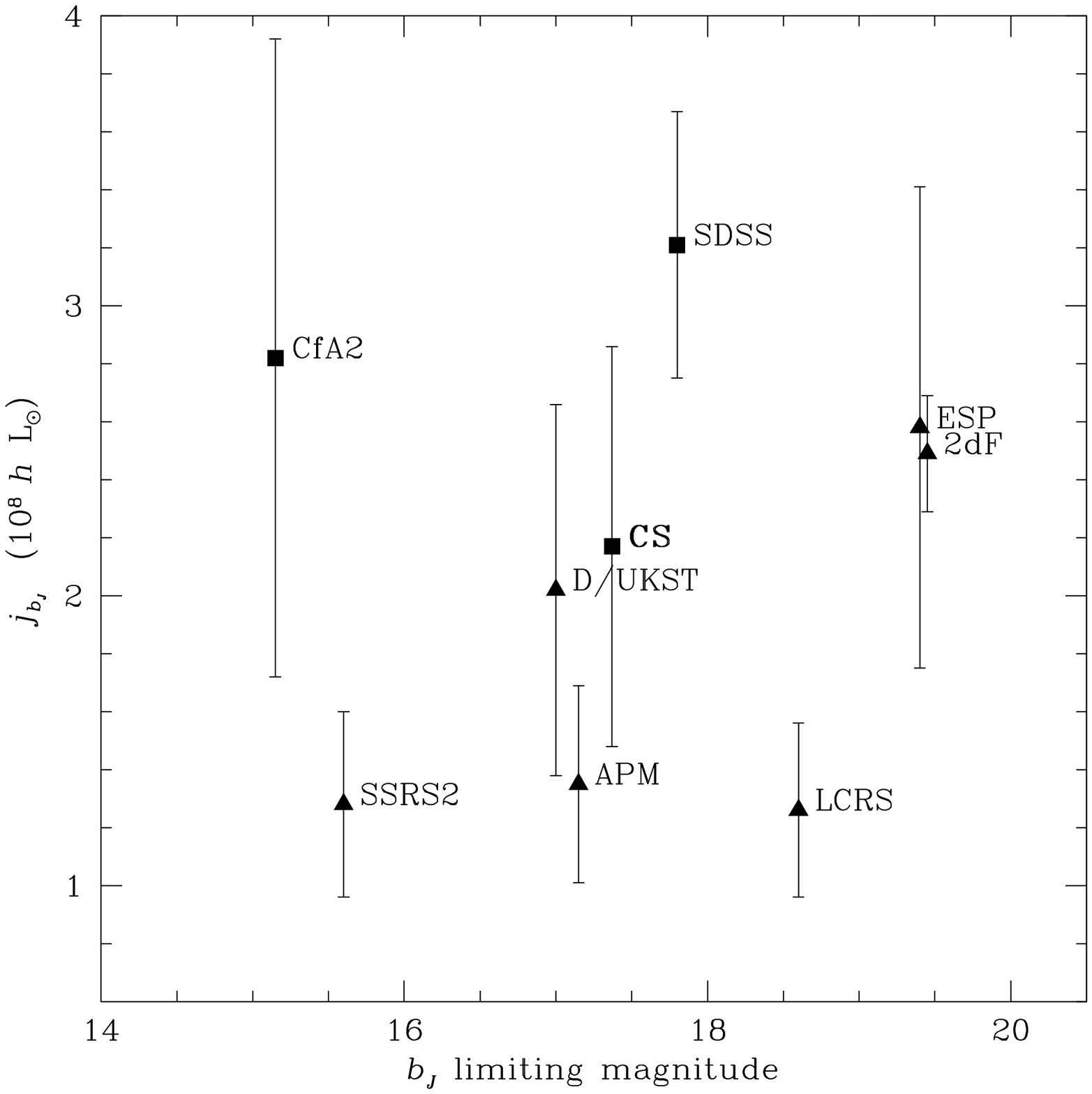}}
\figcaption[wbrown.fig19.ps]{ \label{figj} Luminosity
density $j_b$ and the $b_J$-limiting magnitude of various redshift surveys
(Table \ref{lfs}).  Squares mark the northern surveys, triangles mark the
southern surveys. The CS luminosity density is for the western 64 deg$^2$
studied in this paper; averaging in the eastern half of the CS would raise
the luminosity density to $3.0\pm0.7\times10^8 h L_\sun$ Mpc$^{-3}$.}

\centerline{\includegraphics*[scale=0.5]{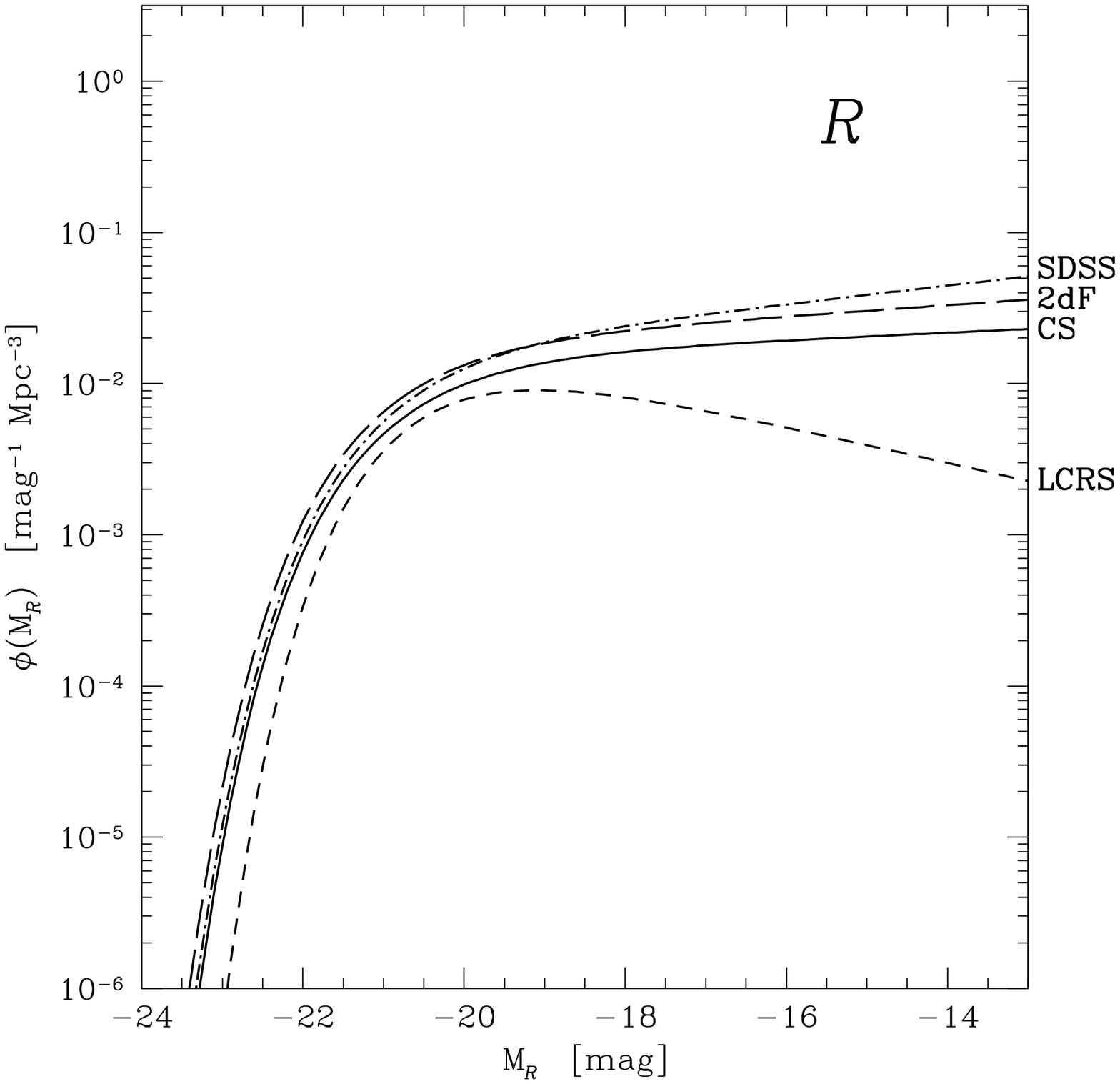}}
\figcaption[wbrown.fig20.ps]{ \label{figall}
Comparison of CCD-based local galaxy luminosity function determinations,
including the current 2dF. We assume $(R-r^*)=-0.22$ for the SDSS
\citep{fukugita96}, $(B-R)=1.2$ for the 2dF \citep{fukugita95}, and
$(R-r)=-0.10$ for the LCRS \citep{shectman96}.}

\clearpage
\appendix
\section{COMPARISON OF CCD AND PHOTOGRAPHIC PHOTOMETRY}

	Our data allows comparison between the photographic and CCD
photometry of 1374 CS galaxies.  The photographic photometry is accurate
and remarkably complete.  CCD photometry offers improvements in precision,
linearity, and resolution over the photographic photometry.

	The CCD photometry reveals a residual non-linearity in the
photographic photometry.  The brightest $R<11$ mag CS galaxies are
typically 0.5 mag brighter than their photographic magnitudes; this
CCD$-$photographic offset decreases with increasing magnitude.  The
original photometry was calibrated by fixing a preliminary zero point on
each plate by comparison to the drift scans of \citet{ramella95} and
\citet{kent93} at $R=16.0$. Data from all plates was then combined and a
preliminary global slope for the scale error was obtained.  The zero point
was then redetermined (at $R=16.0$) for each plate, using the preliminary
relation, and a new global slope was determined.  This process converged
after two iterations.  Our CCD photometry indicates a residual slope of
-0.06 mag / mag --- an error of 10\% in the original zero-point slope.  
This error is consistent with the variation in non-linearity between
photographic plates, and underscores the improvement possible with CCD
photometry.

	Our CCD data, with 0.424 arcsec pixels and FWHM $< 1.5$ arcsec
seeing, has roughly twice the resolution of the scanned photographic
images, with 1.3 arcsec pixels and FWHM $\sim 2$ arcsec seeing.  We
discover 2\% of the 1374 CS galaxies have previously unresolved
companions.  70\% of the resolved objects are companion galaxies.  The CCD
photometry typically places the brightest companion $\sim0.5$ mag fainter
than the original CS ``galaxy.'' 30\% of the close companions are stars
superimposed on faint galaxies; for these objects CCD photometry of the
galaxy typically reveals a $\sim1$ mag offset from the original
photographic catalog.

	Ignoring problematic cases (bright galaxies $R<12$ and galaxies
with previously unresolved companions) the zero-point offset between the
uncorrected CCD and photographic catalogs is $\Delta R_{CCD-photographic}
= 0.014 \pm 0.22$ mag.  Figure \ref{figcscomp} plots the distribution of
$\Delta R_{CCD-photographic}$ for 1314 CS galaxies.  The accuracy of the
CCD and photographic photometry is identical to within the average CCD
zero-point error $\pm0.034$ mag.  The $\pm0.22$ mag scatter, a sum in
quadrature of the CCD and photographic photometric error, shows that the
$\pm0.25$ mag error of \citet{geller97} is slightly overestimated.

	To investigate the CS completeness we apply the original CS
(uncorrected) magnitude limit $R<16.13$ mag and velocity limits $1,000< cz
< 45,000$ km s$^{-1}$ to the CCD and photographic photometric catalogs.
The photographic sample contains $n_{photo}=959$ galaxies, and the CCD
sample contains $n_{CCD}=971$ galaxies. Thus the photographic sample is
$n_{photo}/n_{CCD}=98.8\%$ complete compared to the CCD sample.

	We now explore the properties of the galaxies specific to the
photographic and CCD samples.  The CCD sample excludes 121 galaxies from
the original photographic sample, and includes 133 new galaxies.  60\% of
the 133 galaxies lie within 1$\sigma=0.22$ mag of $R=16.13$.  Thus the
major difference between the CCD and photographic samples is galaxies
scattering above/below the limiting magnitude.  The next most numerous
addition/subtraction of galaxies comes from the 27 previously unresolved
companions.  Of the remaining objects specific to the CCD catalog, only 3
are LSB galaxies.  Using the $\mu(0)_R > 20.8$ mag arcsec$^{-2}$
definition of a LSB galaxy, the photographic catalog contains 12 LSB
galaxies to $R<16.13$ mag and the CCD catalog 15.  Thus the photographic
catalog is 80\% complete compared to the CCD catalog in LSB galaxies. We
conclude that, despite non-linearity and resolution problems, the CS
photographic catalog is nearly as complete as the CCD catalog.

\section{CENTURY SURVEY CCD PHOTOMETRY}

	Table \ref{sample} contains the CCD photometry for the CS $V$ and
$R$ samples.  There are a total of 1295 galaxies.  Column 1 is the galaxy
identification; IDs under 3000 are galaxies listed in the original Century
Survey photographic catalog.  Column 2 is the J2000 right ascension in
hours, minutes, and seconds.  Column 3 is the J2000 declination in
degrees, arcminutes, and arcseconds.  Columns 4 and 5 are the measured $R$
and $V$ magnitudes; the magnitude errors are a sum in quadrature of the
zero-point error and the measurement error.  Column 6 is the $E(\bv)$
extinction value from \citet{schlegel98} for that RA and Dec; we derive
galactic extinctions from $A = E(\bv) \cdot A/E(\bv)$ using $A/E(\bv)$
values from Table 6 of \citet{schlegel98}.  Columns 7 and 8 are $k$
corrections interpolated from \citet{poggianti97} types E, Sa, and Sc and
\citet{frei94} type Im for the Kron Cousins $R$ and Johnson $V$ filters.  
The $V$ and $R$ samples are defined by $V-A_V-k_V<16.70$ and
$R-A_R-k_R<16.20$ mag.

\begin{deluxetable}{lcccccccc}
\tablewidth{0pt}
\tablecaption{Local Galaxy Luminosity Functions\label{lfs}}
\tablecolumns{9}
\tablehead{	\colhead{Survey} & 
		\colhead{$\overline{z}$} &
		\colhead{Area} &
		\colhead{Galaxies} &
		\colhead{Mag Limit} &
		\colhead{$M^*+5\log h$} &
		\colhead{$\alpha$} &
		\colhead{$\phi^*$} &
		\colhead{Ref.} \\
	\colhead{} &
	\colhead{} &
	\colhead{(deg$^2$)} &
	\colhead{} &
	\colhead{} &
	\colhead{(mag)} &
	\colhead{} &
	\colhead{($h^3$ Mpc$^{-3}$)} &
	\colhead{}
}
\startdata
CS & .06 & 102 & 1762 & $R=16.13$ & $-20.73\pm0.18$ & $-1.17\pm0.19$ &
$0.025\pm0.006$ & 1 \\

SDSS & .10 & 140 & 11275 & $r^*=17.6$ & $-20.67\pm0.03$ & $-1.15\pm0.03$ &
$0.019\pm0.002$ & 2 \\

LCRS & .10 & 700 & 18678 & $r\sim17.5$ & $-20.29\pm0.02$ & $-0.70\pm0.05$
& $0.019\pm0.001$ & 3 \\

2dF & .10 & 257 & 20765 & $b_J=19.45$ & $-19.75\pm0.05$ & $-1.09\pm0.03$ &
$0.0202\pm0.0002$ & 4 \\

D/UKST & .05 & $\sim1500$ & $\sim2500$ & $b_J\sim17$ &
$-19.68\pm0.10$ & $-1.04\pm0.08$ & $0.017\pm0.003$ & 5 \\

ESP & .10 & 23 & 3342 & $b_J=19.4$ & $-19.61^{+0.06}_{-0.08}$ &
$-1.22^{+0.06}_{-0.07}$ & $0.020\pm0.004$ & 6 \\

SSRS2 & .025 & 5550 & 5036 & m$_{B(0)}=15.5$ & $-19.43\pm0.06$ &
$-1.12\pm0.05$ & $0.013\pm0.002$ & 7 \\

CfA2 & .025 & 6900 & 9063 & m$_Z=15.5$ & $-18.8\pm0.03$ & $-1.0\pm0.2$ &
$0.04\pm0.01$ & 8 \\

APM & .05 & 4300 & 1658 & $b_J=17.15$ & $-19.15\pm0.13$ & $-0.97\pm0.15$ &
$0.014\pm0.002$ & 9 \\

\enddata
\tablecomments{Local redshift surveys with $>1000$ galaxies and
$\overline{z}>0.01$.  Luminosity function parameters use a $H_0=100$ CDM
cosmology.}
\tablerefs{
(1) \citet{geller97}; 
(2) \citet{blanton01};
(3) \citet{lin96};
(4) \citet{cross00};
(5) \citet{ratcliffe98};
(6) \citet{zucca97};
(7) \citet{marzke98};
(8) \citet{marzke94};
(9) \citet{loveday92}
} 
\end{deluxetable}

\begin{deluxetable}{ccccccc}
\tablewidth{0pt}
\tablecaption{Century Survey LF\label{cs}}
\tablecolumns{7}
\tablehead{	\colhead{$H_0, \Omega_m, \Omega_\Lambda$} &
		\colhead{Band} &
		\colhead{Galaxies} &
		\colhead{$M^*$} &
		\colhead{$\alpha$} &
		\colhead{$\phi^*$} &
		\colhead{$j$} \\
	\colhead{} &
	\colhead{} &
	\colhead{} &
	\colhead{(mag)} &		
	\colhead{} &
	\colhead{(Mpc$^{-3}$)} &	
	\colhead{($10^8 L_\sun$)}	
}
\startdata
100, 1.0, 0.0 & $R$ & 1251 & $-20.84\pm0.09$ & $-1.06\pm0.09$ &
 $0.016\pm0.003$ & $1.88\pm0.59$ \\
& $V$ & 1255 & $-20.20\pm0.09$ & $-1.06\pm0.09$ &
 $0.021\pm0.004$ & $2.21\pm0.69$ \\
75, 0.3, 0.7 & $R$ & 1251 & $-21.57\pm0.09$ & $-1.09\pm0.09$ &
 $0.0060\pm0.0006$ & $1.39\pm0.34$ \\
& $V$ & 1255 & $-20.92\pm0.09$ & $-1.09\pm0.09$ &
 $0.0079\pm0.0009$ & $1.65\pm0.43$ \\
100, 0.3, 0.0 & $R$ & 1251 & $-20.88\pm0.09$ & $-1.07\pm0.09$ &
 $0.016\pm0.003$ & $1.88\pm0.58$ \\
& $V$ & 1255 & $-20.23\pm0.09$ & $-1.07\pm0.10$ &
 $0.020\pm0.003$ & $2.21\pm0.70$ \\
& $R_{\rm blue~1/3}$ & 419 & $-20.51\pm0.17$ & $-1.29\pm0.15$
 & $0.009\pm0.002$ & $0.97\pm0.51$ \\
& $R_{\rm red~1/3}$ & 422 & $-20.69\pm0.14$ & $-0.21\pm0.24$
 & $0.006\pm0.001$ & $0.50\pm0.11$ \\
& $V_{\rm blue~1/3}$ & 415 & $-20.06\pm0.18$ & $-1.43\pm0.15$
 & $0.008\pm0.002$ & $1.05\pm0.59$ \\
& $V_{\rm red~1/3}$ & 424 & $-20.00\pm0.14$ & $-0.13\pm0.25$
 & $0.007\pm0.001$ & $0.56\pm0.12$ \\
\enddata

\end{deluxetable}

\begin{deluxetable}{lccccccc}
\tablewidth{0pt}
\tablecaption{Century Survey CCD Photometry\tablenotemark{a}\label{sample}}
\tablehead{	\colhead{ID} &
		\colhead{RA$_{\rm J2000}$} &
		\colhead{Dec$_{\rm J2000}$} &
		\colhead{$R$} &
		\colhead{$V$} &
		\colhead{$E(\bv)$\tablenotemark{b}} &
		\colhead{$k_R$\tablenotemark{c}} &
		\colhead{$k_V$\tablenotemark{c}}
}
\startdata
cs2     & 8 35 49.11 & 28 53 13.5 & $15.11\pm0.11$ & $15.63\pm0.12$ & 0.050 & 0.02 & 0.04 \\
cs3     & 8 35 54.33 & 29 25 37.9 & $15.93\pm0.04$ & $16.60\pm0.06$ & 0.047 & 0.10 & 0.15 \\ 
cs4     & 8 35 54.60 & 29 41 32.7 & $16.00\pm0.06$ & $16.67\pm0.06$ & 0.053 & 0.14 & 0.20 \\
cs6     & 8 36 03.35 & 28 57 46.9 & $15.53\pm0.04$ & $16.07\pm0.05$ & 0.054 & 0.03 & 0.06 \\
cs8     & 8 36 14.38 & 29 03 16.2 & $15.70\pm0.05$ & $16.36\pm0.07$ & 0.057 & 0.10 & 0.15 \\
\enddata

\tablenotetext{a}{[The complete version of this table is in the electronic
edition of the Journal.  The printed edition contains only a sample.]}

\tablenotetext{b}{Values from \citet{schlegel98}.}

\tablenotetext{c}{Values interpolated from \citet{poggianti97} types E, 
Sa, and Sc and \citet{frei94} type Im for the Kron Cousins $R$ and
Johnson $V$ filters.}

\end{deluxetable}

\end{document}